\documentclass[graybox]{svmult}

\usepackage{type1cm}        % activate if the above 3 fonts are
\usepackage{makeidx}         % allows index generation
\usepackage{graphicx}        % standard LaTeX graphics tool
\usepackage{multicol}        % used for the two-column index
\usepackage[bottom]{footmisc}% places footnotes at page bottom

\usepackage{newtxtext}       
\usepackage[varvw]{newtxmath}       % selects Times Roman as basic font

\makeindex             

\usepackage[skins]{tcolorbox} 
\usepackage{tikz} 
\usetikzlibrary{shapes.geometric, arrows.meta, trees, positioning}
\tcbuselibrary{listings,breakable}
\tikzstyle{arrow} = [thick,->,>=stealth]
\tikzstyle{parent} = [draw, rounded corners=2pt, text width=3cm, align=center, thick] 

\usepackage{pgfplots}
\pgfplotsset{compat=1.18}
\usepackage{subcaption}
\usepackage{nameref}
\usepackage{color,soul} 
\usepackage{booktabs}
\usepackage{siunitx}
\usepackage{lscape}

\newenvironment{takeways}[1][Key Takeaways] 
    {
     \tcbset{colback=red!1!white,fonttitle=\bfseries}
     \begin{tcolorbox}[enhanced,float,floatplacement=h!,breakable,title=#1,frame style={left color=black,
right color=red!75!black}]
     \setlength{\parskip}{0.75em}
    }
    { 
    \end{tcolorbox}
    \setlength{\parskip}{0em}
    }
  
\newenvironment{learningObjectives}[1][Learning Objectives] 
    {
     \tcbset{colback=blue!1!white,fonttitle=\bfseries}
     \begin{tcolorbox}[enhanced,title=#1,frame style={left color=black,
right color=blue!75!black}]
     \setlength{\parskip}{0.75em}
    }
    { 
    \end{tcolorbox}
    \setlength{\parskip}{0em}
    \bigskip
    }

\newenvironment{exercises}[1][Exercise] 
    {
    \tcbset{colback=green!1!white,fonttitle=\bfseries}
     \begin{tcolorbox}[enhanced,title=#1,frame style={left color=black,
right color=green!75!black}]
     \setlength{\parskip}{0.75em}
    }
    { 
    \end{tcolorbox}
    \setlength{\parskip}{0em}
    }

\begin{document}

\title*{Teaching Software Metrology: The Science of Measurement for Software Engineering}

\titlerunning{Theory of Measurement}
\author{Paul Ralph\orcidID{0000-0002-7411-0857} and 
    \\Miikka Kuutila\orcidID{0000-0002-3695-7280} and
    \\ Hera Arif\orcidID{0009-0007-6772-351X} and
    \\ Bimpe Ayoola\orcidID{0000-0002-1802-9819}
}

\authorrunning{Ralph et al.}
\institute{Paul Ralph, Miikka Kuutila, Hera Arif, and Bimpe Ayoola \at  Dalhousie University, Halifax, Nova Scotia, Canada\newline \email{paulralph@dal.ca; miikka.kuutila@dal.ca; hera.arif@dal.ca, bimpe.ayoola@dal.ca}
}
\maketitle

\begin{takeways} 
  \begin{itemize}
   \item Measurement means assigning numbers to phenomena in a principled and systematic way. 
   \item Researchers choose what to measure from webs of causally-related variables
   \item Minimizing random and systematic error decreases the cost, effort, and carbon footprint of research.
   \item \textit{All} measurement is ``theory-laden'' and ``value-laden''; that is, dependent upon multiple concepts, theories, motivations and perspectives.
   \item Multi-item scales or multi-metric instruments and statistical measurement models are required to assess the degree to which instruments measure what they are supposed to measure (construct validity). 
   \item Limited adoption of philosophical realism, multiple measures, and quantitative assessment of reliability and validity, undermines the credibility and effectiveness of software engineering research.  
  \end{itemize}
\end{takeways}

\clearpage 

\abstract{While the methodological rigor of computing research has improved considerably in the past two decades, quantitative software engineering research is hampered by immature measures and inattention to theory. Measurement---the principled assignment of numbers to phenomena---is intrinsically difficult because observation is predicated upon not only theoretical concepts but also the values and perspective of the research. Despite several previous attempts to raise awareness of more sophisticated approaches to measurement and the importance of quantitatively assessing reliability and validity, measurement issues continue to be widely ignored. The reasons are unknown, but differences in typical engineering and computer science graduate training programs (compared to psychology and management, for example) are involved. This chapter therefore reviews key concepts in the science of measurement and applies them to software engineering research. A series of exercises for applying important measurement concepts to the reader's research are included, and a sample dataset for the reader to try some of the statistical procedures mentioned is provided.}

\section{Introduction}
\label{sec:intro}

\begin{learningObjectives}
    \begin{itemize}
        \item Define ``measurement,'' ``measure,'' ``metric,'' and ``instrument''
        \item Give examples of ``measurement,'' ``measure,'' ``metric,'' and ``instrument'' in a software engineering research context
        \item Explain the importance of measurement
        \item Give examples of common measurement problems in software engineering research
    \end{itemize}
\end{learningObjectives}
        
This chapter aims to elucidate key concepts in software measurement, apply them to software engineering research, provide resources for teaching measurement theory that are not otherwise available in this context or in a concise format, and raise awareness of a broad spectrum of measurement issues. 

\textit{Measurement} is the assignment of numerals to objects or events according to rules~\cite{Campbell2013Physics}---often paraphrased as `the principled assignment of numbers to phenomena'---and a \textit{measure} is the number assigned. For example, when we measure the height of a child with a measuring tape, we assign the child a height (i.e. a measure) of, say, 85 cm. This assignment is ``principled'' because:
\begin{itemize}
    \item the numbers on the tape are in the same order as on a base-10 number line,
    \item a cm is a widely understood measure of length that comes from an international standard system of measurement (the International System of Units)
    \item there exists a strong cultural norm (and de facto standard in health research), regarding how children's heights are measured: child standing straight, heels flat on the floor, barefoot, measuring from the floor to the top of the head, etc.
\end{itemize}
In other words, much care, rigor, and effort has gone into creating and validating systems that allow us to assign meaningful heights to children. 

For our purposes, a \textit{metric} is a method, algorithm, or procedure for assigning one or more numbers to a phenomenon. For example, a metric for non-comment lines of code in a Java file might involve counting semicolons. Meanwhile, an \textit{instrument} is a (physical, conceptual, or virtual) tool for measuring something such as a tape measure (physical), a code quality metrics API (virtual), a heart rate monitor (physical and virtual) or a personality questionnaire (conceptual). Question-based instruments are sometimes called \textit{scales}. 
The extent to which an instrument measures what it purports to measure is called \textit{measurement validity}. For instance, a test intended to evaluate intelligence may actually measure of rote learning or memorization skills~\cite{kimberlin2008validity}. In a software engineering context, the science of measurement is sometimes called ``software metrology''~\cite{flater2016rational}. 

Software engineering research---indeed, computing and engineering research more generally---are plagued by measurement problems including: 
\begin{itemize}
    \item Failing to consider that metrics, instruments, scales, etc. may not measure the target property, or may conflate it with other properties (e.g. Github stars conflates quality with popularity).  
    \item Confusing metrics with properties (e.g. assuming that the number of issues in an issue tracker \textit{is} the number of issues with the product).
    \item Failing to evaluate the preciseness and consistency of a metric's results; ignoring measurement error.
    \item Conflating perceived measures with observed measures (e.g. a manager's reported perceptions a developer's productivity may differ substantially from direct observations) .
    \item Assuming measures are ``objective''.
    \item Operationalizing multifaceted variables with individual metrics (mono-method bias).
    \item Selecting problematic metrics; for example, when a program's energy consumption strongly depends on a) whether it uses multi-threading, b) whether it runs on the CPU or GPU; and c) whether it runs on performance cores or efficiency cores, then CPU time is a poor proxy for energy consumption.
    \item Failing to assess, conceptually \textit{and} quantitatively, the validity of a study's measurement strategy.
    \item Assuming that measures are valid because they are popular.
\end{itemize}

Measurement is important because measuring the wrong things, or measuring the right things the wrong way, creates research that superficially seems trustworthy but isn't. Poor measurement makes research more expensive and time-consuming; leads to incorrect conclusions, and undermines the credibility of not only individual researchers but also entire fields of scholarship.

\begin{exercises}

   \textit{Choose a research topic.}
   
    This chapter includes a series of exercises (in boxes like this one) to help the reader apply the concepts to their thesis or research program. 
    
    All subsequent exercises assume the reader has a specific research topic in mind, so the first exercise is to choose a topic. 
    
    If the reader has no idea what research topic to use, they can adopt or adapt one of these examples:

    \begin{itemize}
        \item The impact of immediate feedback for android game developers in Australia on their code's energy consumption.
        \item The causes and consequences of burnout among female software engineers in Sri Lanka.
        \item The effect of AI code generation on the quality of unit tests in open source Java projects. 
    \end{itemize}

    Undergrads should invent topics they personally find interesting, while graduate students should stick with the (perhaps preliminary) topic of their research proposal. 

    Topics should: (1) focus on causal relationships, (2) identify specific groups of subjects or objects, (3) identify one or more specific properties of those subjects or objects, and (4) be statements, not questions. 

    Our suggested exercises are intended for informal use in a flipped classroom, not formal assignments for grading. Exercises can be completed solo but usually its better for students to work in pairs or groups. Students should write down their answers (because writing helps us be more specific) and share with a group or whole class (because explaining things aloud improves sensemaking.)
\end{exercises}

This chapter therefore explores some of the key concepts in software metrology. Understanding these concepts will help researchers envision better measurement strategies, leading to more trustworthy research. Along the way, we provide helpful tips for instructors and exercises for learners. This chapter is accompanied by an online supplement including sample data for students to practice many of the techniques mentioned (see \nameref{sec:Supplements} on page \pageref{sec:Supplements}). 

\begin{exercises}
    \textit{Transform your topic into a research question.}
   
    Here, \textit{research question} denotes a broad concern driving a program of research. Many students will generate poor examples of research questions like: `Does the frequency of Scrum meetings affect perceived intrateam communication?' This is just a hypothesis (`Scrum meeting frequency is directly proportional to perceived intrateam communication') rephrased as a question. Research questions should be simultaneously broader in scope and more specific regarding where and when the phenomenon of interest takes place, and possibly who is involved. A better example of a research question might be: 
    `How do practices associated with Distributed Scrum affect success in software projects conducted anywhere in the world since the advent of Distributed Scrum in the mid-2000s'~\cite{Santos2023Distributed}. 

    Students should share their research questions and discuss how to improve them. 
\end{exercises}

\subsection{Lessons Learned}
The first author has taught research methods for management, computing, and engineering, at multiple levels from undergrad, to junior faculty, including short courses, long courses, and tutorials, on a range of topics from a general introduction to research, to advanced courses in quantitative methods, to courses specifically on measurement. While every offering is unique, one major challenge teaching measurement stands out: getting students to \textit{care} about measurement. Teaching measurement means asking students to fight the system---to eschew the corner-cutting status quo for more rigorous methodology. Construct validity seems abstract and unimportant compared to getting statistical significance or outperforming the latest benchmark. 

To connect with students, instructors need to explain how improving  measurement and construct validity helps address students' core concerns, rather than simply creating new problems. Throughout this chapter, we therefore elevate examples of how better measurement can address students' concerns.

\section{Selecting Variables from the Causal Web} \label{sec:causalWeb}

\begin{learningObjectives}
    \begin{itemize}
        \item Explain what is meant by a causal ``chain'' or ``web''
        \item Draw an example of a causal web 
        \item Analyze the implications of selecting dependent variables from different parts of a causal web
    \end{itemize}
\end{learningObjectives}

Suppose we have built a testing tool that identifies bugs. We believe that identifying bugs is a worthwhile outcome not because it is intrinsically valuable but because of its expected effects downstream. Helping developers find more bugs presumably helps them fix the bugs, which leads to higher software quality, happier users, increased sales and so on. We imagine a causal chain from our dependent variable to more obviously important, higher-level outcomes.

However, most links in our inferred causal chain probably have multiple antecedents (causes) and consequences (effects). User satisfaction, for instance, probably depends on a system's UX design, feature selection, and revenue model as well as bugginess, and affects market share, intention to continue using, the development company's odds of being bought, and the probability that the software will continue to be maintained, among other things. So it is more like a causal web. 

When we design a quantitative software engineering study, we typically choose dependent variables from a causal web leading up to overall software engineering success~\cite{ralph2014dimensions}. Where in the causal web we aim has profound implications for the difficulty and epistemic value of our research. Choosing variables too far away in the web can make some methodological approaches (especially lab-based experiments, quantitative simulations, and benchmarking studies) intractable. But choosing a variable too close in the causal web limits the epistemic value (how much we learn) and practical importance of our work.

\begin{exercises}
    \textit{Look at your research question. Make a list of variables you might want to measure. Include dependent or endogenous variables co-variates, and exogenous variables (but not independent variables).}
    
    Here, we assume the instructor: 
    \begin{enumerate}
        \item has introduced all the variable types mentioned above
        \item will help students determine whether they are manipulating independent variables or measuring exogenous variables
    \end{enumerate}
\end{exercises}

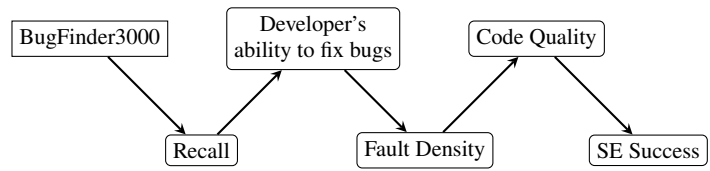
\begin{figure}[t] 
    \centering
    \begin{tikzpicture}[
      node distance=2.1cm,
      start node/.style={shape=rectangle, draw, align=center},
      end node/.style={shape=rectangle, rounded corners=2pt, draw, align=center},
      process node/.style={shape=rectangle, rounded corners=2pt, draw, align=center} 
    ]
    
        \node (start) [start node, above left] {BugFinder3000};
        \node (process2) [process node, below right of=start] {Recall};
        \node (process3) [process node, above right of=process2] {Developer's \\ ability to fix bugs};
        \node (process4) [process node, below right of=process3] {Fault Density};
        \node (process5) [process node, above right of=process4] {Code Quality};
        \node (end) [end node, below right of=process5] {SE Success};
        
        \draw [arrow] (start) -- (process2); 
        \draw [arrow] (process2) -- (process3); 
        \draw [arrow] (process3) -- (process4); 
        \draw [arrow] (process4) -- (process5); 
        \draw [arrow] (process5) -- (end); 
    
    \end{tikzpicture}
    \caption{Simple example of causal chain}
    \label{fig:chain} 
\end{figure}

Back to our bug-finding tool example, suppose we imagine the causal chain shown in Figure~\ref{fig:chain}. We hypothesize that our tool will have better recall than existing tools; that is, it will find more bugs. This will increase software developers' ability to fix the bugs. As developers fix more bugs, the fault density of their systems will decrease, leading to higher code quality, which contributes to overall software engineering success.

\begin{exercises}
    \textit{Look at your list of variables and try to draw a causal chain or web connecting your variables to overall software engineering success (or a similar top-level construct).}

    Students may need a little help identifying all the relevant mediating variables.
\end{exercises}

Selecting ``recall'' as our dependent variable makes our study more tractable. We can get an open source code corpus and test whether our tool can find some bugs that other state-of-the-art tools miss. However, this kind of simulation is essentially a non-experimental design with a single participant: the researcher. We won't know if other software developers will adopt or can even use our tool. Our tool might find bugs that are too hard to fix, have little appreciable effect on code quality, or don't practically matter for the success of the project. Focusing on recall therefore limits our study's epistemic value.

In contrast, selecting overall SE success as our dependent variable decreases feasibility. We simply cannot recruit a representative sample of hundreds of software teams, randomly assign them to a treatment group and a control group, ask the treatment group to use our tool, and measure differences in overall success for, say, the next five years. Therefore, even if it were possible, selecting overall SE success as our dependent variable would make our research too slow and expensive unless we adopt a qualitative approach like action research (see Chapter \hl{XX}: ``Teaching Action Research''). 

The most compelling studies involve dependent variables that are far enough up the web to be ambitious, but not so far as to be impractical. Engaging some human participants to try our bug finding tool, even if those participants are undergraduate students, would have greater epistemic value than just trying it ourselves. 

When a research area is dominated by a de facto standard dependent variables (e.g. recall), it discourages students from imaging higher-level alternatives. In such cases, ask yourself (or your students), \textit{why} do we care about this variable? We might care about recall because better bug finding tools have better recall. This raises the question: what are the other dimensions of quality for this kind of tool? 

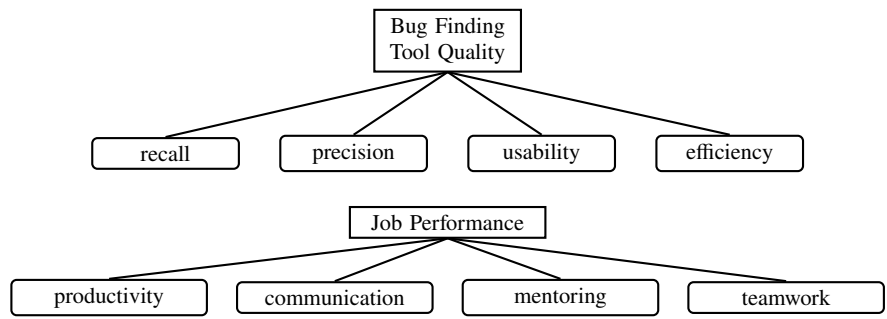
\begin{figure}[t]
    \centering
    \begin{tikzpicture}[
      level distance=1.5cm, 
      sibling distance=2.5cm, 
      parent anchor=south,
      child anchor=north,  
      every node/.style={draw, text width=1.75cm, align=center}, 
    ]
    
    \node {Bug Finding Tool Quality} [parent]
      child {node {recall}}
      child {node {precision}}
      child {node {usability}}
      child {node {efficiency}};
    
    \end{tikzpicture}
    
    \bigskip

    \begin{tikzpicture}[
      level distance=1cm, 
      sibling distance=3cm, 
      parent anchor=south, 
      child anchor=north, 
      every node/.style={draw, text width=2.4cm, align=center}, 
    ]
    
    \node {Job Performance} [parent]
      child {node {productivity}}
      child {node {communication}}
      child {node {mentoring}}
      child {node {teamwork}};
    
    \end{tikzpicture}
    
    \caption{Examples of Abstracting to a Broader Construct of Interest}
    \label{fig:constructDimensions}
\end{figure}

\begin{exercises}
    \textit{If you're having trouble moving beyond a single dependent variable, ask yourself why you care about this variable? What broader construct is this variable a dimension of? Draw a diagram showing the other dimensions of this construct.}

    Students may need some example diagrams (e.g. Fig.~\ref{fig:constructDimensions}).
\end{exercises}

This issue of fixating on a single dimension of a multidimensional construct of interest is not limited to tool testing. Human Factors research in software engineering also tends to fixate on individual dimensions of broader concerns (e.g. productivity is just one dimension of job performance).   

Moving from a single dependent variable to a comprehensive conception of success typically demands more complex, ambitious studies. However, these studies are more likely to produce interesting results that are easier to publish. A new artifact doesn't have to outperform existing artifacts on every possible dimension. Maybe the BugFinder3000 doesn't have better recall, but presents its results in a way that is more helpful to developers, so they're more successful in practice. Maybe your new way of running retrospective meetings doesn't make anyone more productive but improves communication and teamwork. The more quality dimensions you consider, the more likely you'll find something exciting. 

Note: we are not advocating p-hacking. Don't go overboard and select 100 dependent variables. But if the construct you care about has five dimensions, try to measure all five dimensions. 

To summarize, we choose our dependent (or endogenous) variables from a complex, often poorly-understood causal chain (or web) that, we imagine, runs from our independent (or exogenous) variables to high-level concerns like overall software engineering success, human prosperity, or social cohesion. Choosing a dependent variable too close to our independent variable limits epistemic value; choosing a variable too far away makes quantitative, and especially lab-based research, intractable. We need a middle ground. When someone struggles to move beyond too-close dependent variables, ask why we care about this variable? The explanation usually reveals a broader construct of interest, of which the identified variable is just one dimension.

Once we have identified the things we want to measure, we can begin working through how to measure them in a reliable and valid manner.

\section{Assessing and Improving Reliability} 
\label{sec:representational}

\begin{learningObjectives}
    \begin{itemize}
        \item Describe the Representational Theory of Measurement 
        \item Define reliability, random error, systematic error, and measurement invariance
        \item Explain how random error increases the cost of research
        \item Explain how systematic error undermines measurement validity
        \item List techniques for improving reliability
    \end{itemize}
\end{learningObjectives}

The Representational Theory of Measurement (RTM) posits that ``a measurement scale [or metric] is a many-to-one mapping---a homomorphism---from an empirical to a numerical relational structure, and measurement is the construction of scales [or metrics]''~\cite{Tal2020Measurement}. In other words, measurement involves constructing a numerical representation of something. RTM aims to offer a principled and structured approach to convert subjective assessments into objective measurements~\cite{michell1999measurement, Heilmann2015ANI}.

\begin{exercises}
    \textit{Look at the list of variables you want to measure. How can you measure these variables? What instruments, scales, tools, methods, or metrics can you use?}  
\end{exercises}

RTM gives us many concepts for assessing and improving \textit{reliability}. ``Reliability is the consistency of measurement, or stability of measurement over a variety of conditions in which basically the same results should be obtained''~\cite{drost2011validity}. Another way to think about reliability is as a measure of a method's resilience to different kinds of \textit{measurement error}. 

\subsection{Minimizing Measurement Error}
Suppose we want to measure the time it takes for an athlete to complete a 100m sprint. The coach uses a stopwatch, starting the timer when the athlete begins the sprint and stopping it when the athlete crosses the finish line. This approach to measurement may suffer from both random and systematic error.  

\textit{Random error} refers to the variability in measurement results that is caused by unpredictable and uncontrollable factors~\cite{Scott2013}. In our sprint example, physical button-clicking is not perfect. The coach might be off by a $1/10$\textsuperscript{th} of a second or so. 

\begin{exercises}
    \textit{Look at your list of variables. Can you identify any potential random errors in your chosen measurements? }

    If students are adamant that their measures cannot possibly have random errors, see Section \ref{ssec:ReliabilityLessonsLearned}.
\end{exercises}

\textit{Systematic error} refers to a consistent, repeatable error associated with faulty equipment or flawed experimental design. In our sprinting example, suppose the coach always stands half way between the starting line and the finish line. Because of the viewing angle, the coach tends to click start a little late and stop a little early. This will create a systematic error where our observed times are, on average, a little less than actual run times. 

\begin{exercises}
    \textit{What are possible sources of systematic error in your chosen measurements? }

     If students are adamant that their measures cannot possibly have systematic errors, see Section \ref{ssec:ReliabilityLessonsLearned}.
\end{exercises}
 
One way to think about random and systematic error is called \textbf{True Score Theory}. True Score Theory posits that an observed quantity ($X$) consists of the true quantity ($T$) plus or minus some amount of random error ($e_r$) and some amount of systematic error ($e_s$): 

\begin{equation}
    X = T + e_r + e_s
\end{equation}

The difference between the true value and the observed value is the measurement error. 

\begin{figure}[t]
    \pgfmathdeclarefunction{gauss}{2}{%
    \pgfmathparse{1/(#2*sqrt(2*pi))*exp(-((x-#1)^2)/(2*#2^2))}}
    \centering
    \begin{subfigure}{0.49\textwidth}
       \centering
        \begin{tikzpicture}
            \begin{axis}[domain=9.4:10.2,samples=500,yticklabels=\empty,mark=none,width=6.99cm], 
            axis x line*=bottom, 
            axis y line*=left,
            enlargelimits=upper] 
            \addplot+[no markers] {gauss(9.6,0.05)};
            \addplot+[no markers] {gauss(10,0.05)};
            \end{axis}
        \end{tikzpicture}
        \caption{Good reliability}
    \label{fig:goodReliability}
    \end{subfigure}
    \hfill
    \begin{subfigure}{0.49\textwidth}
        \centering
        \begin{tikzpicture}    
              \begin{axis}[domain=9.2:10.7,samples=500,yticklabels=\empty,mark=none,width=6.99cm], 
              axis x line*=bottom, 
              axis y line*=left, 
              enlargelimits=upper] 
              \addplot+[no markers] {gauss(9.9,0.2)};
              \addplot+[no markers] {gauss(10,0.2)};
            \end{axis}
        \end{tikzpicture}
    \caption{Poor reliability}
    \label{fig:badReliability}
    \end{subfigure}
    \caption{Sufficient (a) vs. Insufficient (b) reliability given expected effect size of super shoes (blue) over regular shoes (red)}
\end{figure}
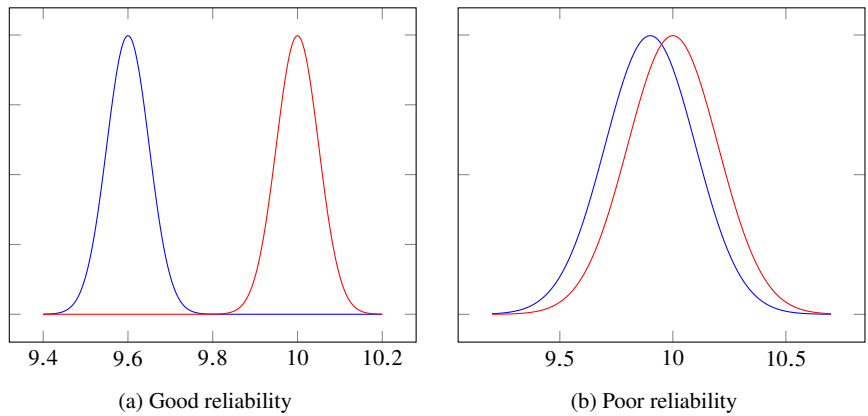

Random error makes your research more expensive and reduces your chances of detecting real effects. Suppose we have invented super running shoes and have recruited some sprinters to test them against regular running shoes. Further suppose that we expect our new running shoes to shave $0.4$ seconds off a 100m sprint time, and our stopwatch-wielding judge has a standard random error of $0.05$ seconds. Figure~\ref{fig:goodReliability} shows the probability distribution of our judge's likely observations. That is, if the true score is 9.6 seconds, our judge will probably record an observation between 9.5 and 9.7 seconds, and if the true score is 10 seconds, our judge will probably record an observation between 9.8 and 10.1 seconds. These intervals don't overlap, so we should be able to detect our expected effect. 

But what if our new shoes are only 0.1 seconds faster, and our human-with-a-stopwatch has a higher standard error---say 0.2 seconds. Figure~\ref{fig:badReliability} shows the probability distribution of our judge's likely observations. See all the overlap? Now, due entirely to random error, we might \textit{observe} that our super shoes are \textit{slower} than regular shoes, even though the opposite is true. 

In situations where random error is large compared to expected effect size, we either need to run heaps of trials and average the results, or employ numerous judges and average their scores, or both. With enough observations, the random errors cancel each other out, and we get more accurate estimates of the true score (the well-known ``law of large numbers''). However, consider the less known (because we just made it up) ``law of bad measurement is expensive'': more judges + more participants = more time + more money. While this example is relatively simple, the same principle applies to more sophisticated statistical approaches. 

Systematic error, unlike random error, biases results in one direction leading to consistent and repeated inaccuracies. Suppose, instead of a sprinter, we have a marathon runner. Our runner uses a special watch to measure their heart rate, and follows a sophisticated running program where they have to maintain a certain heart rate for a certain duration. Now suppose that the heart rate monitor consistently underestimates true heart rate by 10 beats per minute (bpm). This could sabotage their whole training program. The runner will constantly push too hard, and won't understand why they can't maintain the target heart rate for the expected duration. 

\begin{figure}[t]
    \pgfmathdeclarefunction{gauss}{2}{%
    \pgfmathparse{1/(#2*sqrt(2*pi))*exp(-((x-#1)^2)/(2*#2^2))}}
    \centering
        \begin{tikzpicture}
            \begin{axis}[domain=80:150,samples=100,yticklabels=\empty,xtick={80,90,100,110,120,130,140,150}, mark=none,width=\textwidth,height = 6cm], 
            axis x line*=bottom, 
            axis y line*=left,
            enlargelimits=upper] 
            \addplot+[no markers, color=blue] {gauss(120,5)}; 
            \addplot+[no markers, color=red] {gauss(110,5)}; 
            \end{axis}
        \end{tikzpicture}
        \caption{Systematic Error in Heart Rate Monitor. True heart rate (blue) vs. Measured heart rate (red), underestimating by 10 BPM}
    \label{fig:systematicError}
\end{figure}
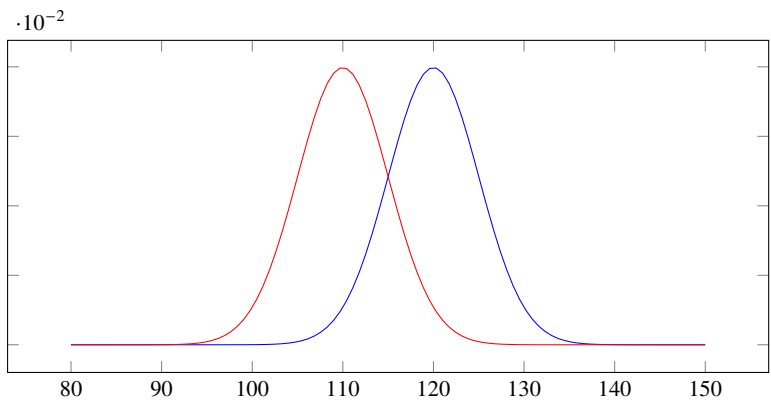

Figure~\ref{fig:systematicError}, illustrates the probability distributions of the true and measured heart rates. The heart rate monitor persistently underestimates the true heart rate by 10 beats per minute. As a result, we might \textit{conclude} that the runner’s heart rate is consistently lower than it actually is, even though this isn't the case.

Systematic error can cause numerous problems, including: 
\begin{itemize}
        \item Measurements that are consistently too high or too low compared to the true value.
        \item Drawing incorrect conclusions from seemingly consistent patterns.
        \item Using more complicated (i.e., harder to understand and explain; easier to mix up; sometimes less sensitive) statistics. Addressing systematic error often requires more advanced statistical techniques such as mixed effects models or structural equation modeling, which are harder to interpret, more prone to being misapplied, and sometimes less sensitive.
    \end{itemize}

\subsection{Measurement Invariance} \label{ssec:MeasurementInvariance}

Continuing with our marathon runner heart rate example, suppose that the reliability of our heart rate sensor is inversely correlated with heart rate: the higher the heart rate, the less accurate the readings~\cite{martin2023activity}. In other words, measurement \textit{varies} across the range of the phenomenon being measured. 

What we want is measurement \textit{in}variance. That is, we want our instruments to measure the same thing, in the same way, across experimental groups or the range of the phenomenon of interest. If we're comparing performance of our running shoes at different running speeds, but our performance measurement (e.g. heart rate) works differently at different speeds, it can distort our results. Similarly, if we use a questionnaire to investigate gender (or cultural, or age-related, etc.) differences in the perceived comfort of our shoes, and the questions have different meanings to people of different genders (cultures, ages, etc.) it can mess up our results. 

A classic example of this is the measurement of childhood depression. In a longitudinal study following children into early adulthood, we might have a series of questions to agree-disagree items like ``the child cries easily''. A 19-year-old male who cries easily is more likely suffering from depression than a 6-year-old female who cries easily. 

\begin{exercises}
    \textit{Could any of your chosen measurements exhibit measurement variance? How?}
\end{exercises}

A comprehensive review of statistical approaches for assessing measurement variance is beyond the scope of this chapter. Here, we simply want to encourage researchers to reflect on their instruments and to question their measurement invariance. Measurement invariance is especially important in longitudinal research (where the meaning or accuracy of a measurement changes over time), questionnaire-based research (where questions have different meanings for different people), and research involving biometrics (where many instruments are only reliable under specific conditions). For more information about establishing measurement invariance, see Putnick and Bornstein~\cite{putnick2016measurement}.   

\subsection{Types of Reliability}

There are three main types of reliability: inter-rater, test-retest, and internal consistency. 

\subsubsection{Inter-rater Reliability and Agreement} \label{ssec:IRR}

Inter-rater (AKA inter-judge, inter-reviewer) reliability is the extent to which different (typically human) observers give \textit{consistent} estimates or ratings~\cite{gwet2014handbook}. Inter-Rater Agreement is the extent to which different (typically human) observers give \textit{identical} estimates or ratings.

In our sprinting examples above, we wouldn't expect multiple coaches with stopwatches to give identical estimates in milliseconds. Therefore, we might assess their ``inter-coach'' reliability using Cronbach's alpha, a statistic used to evaluate the reliability of a test based on the average correlation among items within the test~\cite{tavakol2011making}. We would expect good reliability (e.g. $\alpha>0.8$), not merely statistical significance. 

While alpha is very popular, however, it makes several problematic assumptions~\cite{agbo2010cronbach} that newer procedures, including McDonald's Omega and Composite Reliability, overcome~\cite{zinbarg2005cronbach, cerri2023insomnia}. Furthermore, if our raters were classifying data into nominal categories (e.g. brand of running shoes) or ordinal categories (e.g. self-reported running ability: beginner, intermediate, advanced), we would expect identical ratings and a measure of inter-rater agreement would be more appropriate. 

We assess inter-rater agreement with adjusted statistics, such as Krippendorff's Alpha~\cite{krippendorff2018content}, and unadjusted statistics such as percent agreement. Krippendorff's Alpha is generally preferred; however, there exist situations such as imbalanced data where it will produce nonsense and percent agreement is superior. 

In software engineering, we frequently see inter-rater agreement in the context of systematic reviews, where multiple raters must apply selection criteria. 

\subsubsection{Test-retest Reliability} 

Test-retest reliability refers to the stability of a test from one measurement session to another using the same sample. Software engineering professionals often encounter test-retest reliability issues. Suppose we run an automated test, like a unit test, on some code, and the test passes. Then, without making any changes to the code, we run the test again, and it fails. Programmers call these unstable tests ``flaky;'' scientists call them ``unreliable.'' Poor test-retest reliability often results from random error. 

\subsubsection{Internal Consistency}

Internal consistency arises when we use multiple methods to assess the same property. It is the extent to which all the items in a multi-item scale measure the same behavior or characteristic~\cite{tavakol2011making}. Several coaches with stopwatches are not ``multiple methods'', but one method applied by multiple people. Multiple methods is more like measuring sprint time using a person with a stopwatch, a transponder attached to the runner, and lasers at the start and finish. This might be a bit silly in the context of sprinting but using multiple methods alleviates systematic error. 

Internal consistency is often assessed using Cronbach's alpha.   

\subsection{Improving Reliability}

Techniques for improving reliability include the following.

\begin{itemize}
    \item \textbf{Pilot test instruments.} Administering instruments on a small sample helps to evaluate the feasibility and appropriateness of research instruments and methods, and to identify and fix practical or technical issues, before conducting a major study.

    \item \textbf{Train assistants thoroughly.} In studies involving multiple research assistants making observations, developing a standard protocol and rigorously training assistants helps improve uniformity in data collection and measurement.  

    \item \textbf{Use scripts rather than entering, cleaning, or transforming data manually).} Manually entering or transforming data is error-prone; automated scripts improve accuracy, traceability, auditability, and our ability to fix mistakes quickly.

    \item \textbf{Double-entering data.} Data that must be entered or labeled manually should be independently entered and labeled by at least two researchers. The resulting datasets can then be compared to resolve discrepancies. Reconciliation should be documented carefully and automated where possible. Manual labeling should be performed iteratively, calculating reliability after each round. As disagreements are resolved, researchers should clarify their labeling rules to avoid similar disagreements in subsequent rounds.  
    
    \item \textbf{Statistical procedures for modeling error (mixed and random effects models).} Fixed, random, or mixed-effect models can be used to account for systematic errors, random errors, or both, respectively. 
    
    \item \textbf{Multiple measures and Triangulation.} Using multiple measures can enhance reliability by allowing for data triangulation (see Section~\ref{sec:modelbased}). 
\end{itemize}

\begin{exercises}
    \textit{How can you improve the reliability of your chosen measurements?}
\end{exercises} 

\subsection{Lessons Learned} \label{ssec:ReliabilityLessonsLearned}

Computer science and engineering students tend to grasp easily the ideas of reliability and true score theory, and how unreliable measures increase research cost and effort. However, they often struggle with the possibility that \textit{their} (computerized, mostly deterministic) instruments are unreliable. The near-perfect test-retest reliability of \textit{some} deterministic software running on the same inputs in carefully controlled laboratory settings creates the illusion of reliability. Discussing diverse examples can help. For instance:
\begin{itemize}
    \item At the time of writing, research involving Large Language Models (LLMs) is in vogue. As long as LLM output is nondeterministic, any measurement process using them is unreliable.
    \item One student gave a great example of measurement invariance when measuring soil conditions using sensors that connect to satellites. The cloudier the sky, the more likely data is missing, leading to systemic bias in temperature and moisture readings 
    \item Reviewing causes of flaky tests (see~\cite{Parry2021Flaky}) is a great opportunity to discuss a measurement issue that's highly relevant to industry, and emphasizes the fact that ostensibly deterministic computerized measurement can still be unreliable.
\end{itemize}

In contrast, some researchers have very simple instruments; for instance, measuring the quality of open source projects based on the number of stars they have on Github. This example, which has great reliability but terrible validity, makes an excellent segue to our next topic. 

\section{Understanding Validity from a Realist Perspective} 
\label{sec:realism}

\begin{learningObjectives}
    \begin{itemize}
        \item Explain in broad strokes how realism differs from positivism and interpretivism. 
        \item Define and give examples of \textit{theory-laden} and \textit{value-laden}
        \item Define \textit{generative mechanism}
        \item Explain how to decide whether to \textit{treat} a variable as latent 
        \item Give examples of strategies for coping with the theory-laden and value-ladenness of observation
    \end{itemize}
\end{learningObjectives}

The previous section focused on reliability---the degree to which measures are consistent. But what if you reliably measure the wrong thing? To understand \textit{validity}---the degree to which an instrument measures what it is supposed to measure---we need to discuss a little philosophy. Unfortunately, most computer scientists don't get much exposure to philosophy of science, and when they do, they they tend to learn about outdated epistemological approaches like positivism and falsificationism. However, the whole idea of validity is rooted in philosophical \textit{realism}, not \textit{positivism}.

\subsection{The Problem of Induction}

Usually, the whole point of writing a scientific article is to make some claim about the world like: `The new BugFinder3000 finds more bugs than the old BugFinder2000', `Distributed Scrum has no impact on project success''~\cite{Santos2023Distributed}, or `the four dimensions of software sustainability are environmental, social, economic and technical'~\cite{McGuire2023Sustainability}. The fundamental question in the philosophy of knowledge (epistemology) is how we can justify such claims about the world. The key problem with such justification was formulated by Hume:  

\begin{quote}
    Thus, not only our reason fails us in the discovery of the ultimate connexion of causes and effects, but even after experience has inform’d us of their constant conjunction, ’tis impossible for us to satisfy ourselves by our reason, why we shou’d extend that experience beyond those particular instances, which have fallen under our observation.~\cite{hume1739treatise}
\end{quote}

In other words, suppose that every time we observe a software team transition from Waterfall to Agile, their project succeeds. We want to claim that the Waterfall-Agile transition \textit{causes} success. But how do we know that the pattern we see in the transitions we observed also holds in transitions we did not observe (e.g. elsewhere, under different conditions, or in the future)? This is called the \textit{Problem of Induction} because Hume questions the inductive leap from `we observe this pattern in instances A, B, and C' to `this pattern exists in unobserved (e.g., future) instances X, Y, and Z'. Put another way, Hume questions how we figure out whether nature is uniform across space and time. The various epistemological schools that emerged in the 20\textsuperscript{th} century can be differentiated according to their approach to Hume's challenge. 

\subsection{From Positivism to Interpretivism}

Positivists (or more properly, ``Logical Empiricists'') such as Rudolph Carnap argued that we justify scientific claims by finding supporting empirical observations. The more observations we have confirming the pattern, the greater the probability that the pattern will hold in the future. In hindsight, this doesn't address Hume's challenge at all. Over time, the positivists realized that no amount of past observations conforming to a pattern could prove the pattern would continue to hold, and many of them abandoned positivism as unworkable. 

Falsificationists, led by Karl Popper, accept that induction is not justified but deny that science is inductive. They view scientists as searching for observations that refute (or ``falsify'') causal claims.  This doesn't address Hume's challenge either. For a single observation to refute a theory ``it must be presupposed that the course of nature will not change so that the experimental and observational context in which the refuting observation statement is true ceases to be true''~\cite[p. 91]{archer2013critical}. In other words, falsificationism assumes that nature is uniform---the very thing Hume questioned.

Failing to address Hume's Problem of Induction isn't the only issue with Positivism and Falsificationism. Back then, philosophers thought about cause and effect as ``constant conjuction'' (as in Hume's quote above); they thought $x$ causes $y$ if and only if, whenever $x$ occurs, $y$ follows. Most scientists now think of causality probabilistically: $x$ causes $y$ if $p(y|x)>p(y|!x)$. Positivists and Falsificationists also thought that individual theories could be isolated for testing. But, as argued by Imre Lakatos, science is characterized by constellations of interconnected theories such that no one theory can be tested (and confirmed or falsified) in isolation. Furthermore, when an observation appears to confirm (or refute) a theory, it may be because the study was flawed, the observation was recorded incorrectly, the instruments were unreliable, the math connecting the observation to the theory was done wrong, or the expected observation was derived incorrectly from the theory. Lots of things can go wrong, and the same thing can go wrong over and over, so no number of confirmatory or disconfirmatory observations can verify or falsify a theory. (This is called the Quine-Duhem Thesis.) 

Crucially, positivists and falsificationist assumed that observation was unproblematic and measures were inherently valid. Construct validity is not, and never has been, a positivist quality criterion; the whole idea of construct validity is rooted in realism (below). 

Positivism is as dead as a philosophical movement can be~\cite{passmore1967logical}. Falsification is, similarly, an epistemological cul-de-sac because its proponents have utterly failed to address the conceptual challenges presented to it.  

Many software engineering researchers are instead attracted to pragmatism---the philosophical movement associated with Charles Sanders Peirce, William James, and John Dewey. Pragmatists basically argue that truth is inextricable from usefulness---we should judge theories (and measurement instruments) based on how they enable social progress. Contemporary pragmatists might argue that induction is justified \textit{because it is useful}, regardless of the uniformity of nature, and the realist Bhaskar makes an analogous argument (see below). But if nature isn't uniform, a theory that appears useless in one place and time might be useful in another, and vice versa.  Regardless, pragmatism is unhelpful for software metrology because pragmatists would argue that measurement reliability and validity are secondary to usefulness, \textit{thus presupposing reliable and valid measures of usefulness} without advancing substantive approaches to reliability or validity. 

Moving on, Interpretivists (and Postmodernists) respond to Hume by agreeing that induction isn't justifiable; therefore, scientists should focus on understanding the meaning people ascribe to their experiences instead of searching for universal laws. Interpretivists tend to prefer qualitative research methods, generate detailed accounts of specific events without generalizing to other events (past or future), and fashion their interpretations of events into internally-coherent concepts and theories. Interpretivism makes sense, but is deeply unsatisfying. If we want to know how a specific group of software professionals \textit{feel} about something (e.g. generative AI, burnout, sustainability, mutation testing), Interpretivism is great. But it doesn't help us compare multiple competing technologies (or practices or theories). From an Interpetivist perspective, your account and my account of the same events can be completely different and yet equally valid as long as they are both internally coherent, regardless of whether either account corresponds to real events, people, or objects. 

\subsection{Realism}

This brings us to \textit{Scientific Realism}---the view that unobservable structures and processes postulated by science exist in the real world, whether or not humans exist to imagine them---and \textit{Critical Realism}, Roy Bhaskar's sweeping philosophical project that aims to address Hume's Problem of Induction by charting a course between positivism and interpretivism. Realism entails many new ideas that are essential to contemporary approaches to measurement validity. 

First, Hume Carnap, Popper, and their contemporaries thought  causality exists \textit{in our minds}. Realism, in contrast, posits that causality exists in the real world, independent of human observers. Objects have powers and can influence each other (e.g. the sun has the power to warm the Earth, software has the power to frustrate users). 

Second, echoing the inverted thinking of pragmatism mentioned above, Bhaskar argues that nature must be \textit{somewhat} uniform because science is so successful. If physical and social reality was totally unstable, we wouldn't be able to successfully land a dune buggy on Mars, treat skin cancer, reign in inflation, or improve kids' reading skills. Obviously science and engineering are not \textit{always} successful and some things change, but nature is \textit{somewhat} uniform, especially in the short term. 

Realists argue that our assertions of causal relationships are justified by two things: the magnitude and quality of the body of empirical evidence supporting the relationship, and the degree to which we understand the \textit{generative mechanism}. The generative mechanism is \textit{how} $x$ causes $y$. Suppose we want to claim that the degree of formality in the presentation of software desiderata ($x$) reduces design creativity ($y$). We conduct an experiment that shows that $x$ and $y$ are inversely related, shows that $x$ precedes $y$, and controls for third variable explanations~\cite{Mohanani2021Requirements}. Realists argue that, to justify our claim, we still must explain \textit{how} desiderata presentation affects creativity. We need a different kind of study: in this case, one that shows that the more formal desiderata presentation discourages critical thinking~\cite{Mohanani2022Templated}. Because we understand \textit{how} $x$ causes $y$, we can be more confident the relationship will endure. 

Third, realists believe that many of these generative mechanisms (e.g. critical thinking) cannot be directly observed. Indeed, realists believe that reality is full of (``latent'') structures that produce observable \textit{effects}, but cannot be observed directly. These structures exist in both natural science (e.g. quasars, dark energy, the Earth's mantle) and social science (e.g. culture, socioeconomic class). 

\subsection{Latent Variables}

Recall that Positivists assumed observation was unproblematic. Want to know how many ducks are in the pond? Count them. Sure you might miss a couple or count the same one twice, but it's not that complicated. Compare counting ducks to assessing the morale of a software team. You can't see morale. You can make up some survey questions that \textit{aim} to measure morale but how do you know they don't accidentally measure something else, like wellbeing, or conflate morale with other factors? 

Quantifiable properties of latent processes and structures, like morale, are called \textit{latent variables} or \textit{constructs}. Since we cannot observe constructs directly, we must estimate them from variables we \textit{can} observe (see Chapter \hl{XX}: ``Theorizing in Software Engineering Research'').

\begin{exercises}
    \textit{Look at your list of variables. Which are latent? Which are directly measurable? Are you sure?}

    Students may need help disentangling latent variables from common proxy measures. If you can't figure it out, say it's a good example of a tough call, and come back to it in the next exercise.  
\end{exercises}

Scientists have created many methodological and statistical approaches for investigating the validity of instruments for measuring latent variables (see Section~\ref{sec:modelbased}). For now, the key point is that differentiating between less problematic observations (regular variables) and more problematic observations (latent variables) is important because whenever problematic observations demand construct validity assessment (see Section~\ref{sec:modelbased}). 

\textit{Statistically} variables are either latent (problematic) or not (unproblematic), but some realists (particularly Sayer~\cite{sayer2010method}) argue it's more of a spectrum. \textit{All} measurement is \textit{theory-laden}; that is, predicated upon theoretical assumptions made by researchers when designing their studies and choosing what and how to measure~\cite{tal2013old, kuhn1961function}. For example, to count the bugs found by the BugFinder3000 we need a theory of what is and is not a bug. Herzig et al.~\cite{Herzig2013bug} defined a \textit{bug} as a ``request for corrective code maintenance,'' but the IEEE defines a bug as a(n): 
\begin{quote}
    \begin{enumerate}
        \item ``manifestation of an error in software'',
        \item ``incorrect step, process, or data definition in a computer program'',
        \item ``situation that can cause errors to occur in an object'' or
        \item ``defect in a system or a representation of a system that if executed/activated could potentially result in an error''~\cite[p. 179--180]{IEEE2017Vocab}.
    \end{enumerate}
\end{quote}

\noindent None of the IEEE standard definitions have anything to do with requesting corrective code maintenance. Researchers with different theories of bugs may disagree about how many bugs the BugFinder3000 finds. Even individual \textit{concepts} come with their own assumptions or expectations. Our measurement of bugs is predicated on myriad theoretical concepts including software, source code, software behavior, errors, users, programmers, and expectations. If two people have different concepts of ``user'', for example, they might disagree on the nature of bugs. 

\begin{exercises}
\textit{Look at your list of things to measure. What theories and concepts underlie your research? If no theory is obvious, ask yourself, if this were an exam question and you \textbf{had} to identify a related theory, what theory would you choose?}
\end{exercises}

Similarly, all measurement is \textit{value-laden}~\cite{sayer2010method}; that is, predicated upon the motivations and justifications of researchers~\cite{ward2021value}---basically what we think is important. For example, users and developers often disagree about what is and is not a bug. When our statistical analysis package gives an error message that seems clear to the developer but confusing to the user, the former might claim it's desired behavior while the latter claims it's a usability bug. Much software engineering research ignores these value conflicts. For example, when Herzig et al.~\cite{Herzig2013bug} suggested guidelines for classifying issue reports as bugs vs. non-bugs, they totally sidestepped the issue of who, exactly, agrees whether the reported behavior is desirable or not. Just because someone dislikes a behavior enough to write a bug report doesn't mean all relevant stakeholders will agree that the behavior should be changed. 

\begin{exercises}
\textit{Look at your list of things to measure. What values underlie your research? Whose perspective are you adopting? Who might disagree with your conceptualizations?} 

People tend to underestimate how different other reasonable people's perspectives can be. This exercise is best done in pairs of students with different research areas, where one plays devil's advocate for the other. 
\end{exercises}

So it's not so much that all variables are either directly observable or latent. It's more like, all observation is predicated upon certain theories, concepts, and values. The less confident we are in the theory or the less people agree on the values involved, the more problematic the variable, and the more important it is to treat the variable as latent.

For example, when a study participant gives their age as 28 years, we're pretty confident in the underlying theories and values (the Gregorian calendar, meaning of a year, etc.). We have widespread agreement within the scientific community on the meaning of statements like ``the mean age of participants was 31 years.'' Therefore, we treat age as a direct measurement, and assume it is \textit{inherently valid}. 

In contrast, the performance of our BugFinder3000 is predicated on many theories in which we should not be nearly so confident, and many values upon which people disagree. Since theory development in software engineering research is lacking~\cite{johnson2012s,stol2015theory}, most of our measures are problematic. The trouble is that when you are immersed in the norms and perspectives of a specific research area, it's easy to overestimate consensus. Many articles simply count bugs as if everyone agrees on what a bug is. Discussing your ideas with people from very different academic backgrounds (e.g. sociologists, historians, psychologists, microbiologists) can help surface you implicit assumptions. Adopting an explicit theory to guide your work also helps. 

Because all observation is theory- and value-laden, theory is essential for research. Without theory, construct validity is meaningless and researchers implicitly assert (with no evidence) that everyone agrees with their perspective and all of their measures are intrinsically valid. For more on validity threats from the critical realist perspective, see~\cite{johnston2010critical}.

\subsection{Summary}

To summarize, all observation is problematic, but some measurements are more problematic than others. Realism challenges us to address the theory- and value-laden nature of measurement by treating more problematic variables as latent. This gives rise to construct validity---the degree to which instruments measure what we intend them to measure. Researchers doing predominately quantitative studies cannot sidestep construct validity concerns by simply claiming to be positivists, falsificationists, or pragmatists.\footnote{Qualitative researchers \textit{can} dismiss construct validity concerns by embracing interpretivism, but claiming to be an interpretivist means you're not doing predominately quantitative research.} It is indefensible to simply assuming that all our measures are unproblematic without any coherent response to the thesis that all observation is theory- and value-laden. 

\subsection{Lessons Learned}

Teaching realism is quite challenging. Critical realism is a huge philosophical project with a whole mess of novel concepts and jargon. Its seminal works presuppose extensive knowledge of philosophy of science and Bhaskar, in particular, is as impenetrable as he is profound. Furthermore, some of critical realism's key implications are inconvenient for researchers and cast doubt on large swaths of computer science and engineering research. Sayer's argument that \textit{all} measurement is theory- and value-laden cannot be sidestepped by claiming that a study is positivist rather than realist. But students may not want to rock the field's dominant positivist boat, let alone sink it. 

To cope, we recommend (1) sticking to the most important and relatable concepts, as described above, while perhaps sidestepping some of the more esoteric points; and (2) focusing on modest steps toward realism;  for example, adjusting our methods better to understand and explain \textit{how} hypothesized causal relationships manifest (i.e. generative mechanisms). Most reviewers will not be offended when a paper seeks to understand why a relationship is observed. Similarly, one can consider and specify the concepts underlying their measurement strategy without directly claiming that everyone else's measurement is irreparably broken.

Many students struggle to identify the concepts and theories underlying their work (see Section \ref{sec:realism}). This is often due to a lack of immersion in relevant psychological and sociological theory. There is no easy remedy for this. At the graduate level, students and graduate programs just need to prioritize reading. 

\section{Model-based Theory of Measurement}
\label{sec:modelbased}

\begin{learningObjectives}
    \begin{itemize}
        \item Describe model-based approaches to measurement and statistical measurement models
        \item Explain why measurement models are needed to assess technical properties of software systems 
        \item Explain the difference between formative and reflective measurement models, with examples
        \item Describe the general strategy for evaluating a set of reflective indicators 
    \end{itemize}
\end{learningObjectives}

\subsection{Statistical Measurement Models}

Embracing the reality that observation is theory- and value-laden means that:
\begin{enumerate}
    \item for every property we wish to measure, we must decide whether to treat it as unproblematic or latent; and
    \item for every property we treat as latent, we must \textit{investigate} the validity of our instruments, scales, or metrics. 
\end{enumerate}

\noindent Not consider. Not discuss. \textit{Investigate}. 

To investigate construct validity we need a \textit{measurement model} in which each latent property is operationalized as the shared variance of multiple indicators. As Tal explains: ``According to model-based accounts, measurement consists of two levels: (i) a concrete process involving interactions between an object of interest, an instrument, and the environment; and (ii) a theoretical and/or statistical model of that process, where ``model'' denotes an abstract and local representation constructed from simplifying assumptions. The central goal of measurement according to this view is to assign values to one or more parameters of interest in the model in a manner that satisfies certain epistemic desiderata, in particular coherence and consistency''~\cite{Tal2020Measurement}. 

In other words, a model-based approach to measurement acknowledges the existence of latent structures and processes, and proposes a way to measure them by theorizing relationships between latent  variables (more problematic) and indicators thereof (less problematic). We need multiple indicators for each latent variable because the degree to which the indicators of $X$ converge with each other and diverge from indicators of other latent variables provides evidence that the indicators do, in fact, measure what they're supposed to measure.

\subsection{Technical Properties of Software Systems are \textit{all} Latent}

In software engineering research, we usually see statistical measurement models used with questionnaire surveys. Job satisfaction, for example, might be measured using a questionnaire comprising MacDonald and Maclntyre's 10-item generic job satisfaction scale~\cite{macdonald1997generic}. We rarely see statistical measurement models in repository mining, benchmarking, or other lab-based quantitative research. (Using machine learning to predict a variable based on some training data is not a measurement model). 

However, most technical properties of software systems are just as problematic (theory- and value-laden) as psychosocial phenomena like job satisfaction. 

Consider the size of a software system. Size metrics are predicated on numerous theoretical concepts such as source code, code ``lines'', functions, methods, classes, packages, libraries, function points, etc. Size metrics are also predicated upon value judgments. For instance, when measuring the size of software system, should we include the programming language's standard library or third-party libraries? What about services or microservices upon which the system depends? Size is value-laden because people with heterogeneous perspectives, contexts, and goals might reasonably answer these questions differently.

Behavioral properties of software systems (e.g. efficiency, responsiveness) and common code quality dimensions (e.g. understandability, maintainability) are similarly problematic. So are code smells (code characteristics often indicative of underlying problems). Consider the \textsc{long method} code smell. It is predicated on theoretical concepts including source code, method, code line, etc. Furthermore, we can unambiguously define any method exceeding 80 non-comment lines of code as ``long,'' but the threshold selection and exclusion of comments are value-laden choices. And even if everyone agreed, we remain faced value judgments about whether each long method actually is or indicates a problem. 

No one is advancing a substantive argument for treating properties of software systems as unproblematic. While previous articles have advocated for more sophisticated measurement approaches (e.g. \cite{ralph2018construct,graziotin2021psychometrics,russo2021pls}), many individual studies continue to ignore construct validity. 

Acknowledging the theory- and value-laden nature of observations of not only psycho-social phenomena but also technical properties of software systems clarifies the need for a model-based approach to measurement. Now, we can begin taking construct validity seriously using measurement models.

\subsection{Types of Measurement Models}

Approaches to measurement modeling include common factors~\cite{basilevsky2009statistical}, projection to latent structures (PLS) using partial least squares regression~\cite{hair2019pls}, and forged concepts~\cite{Liu2022Forged}. All three of these approaches can be used with multidimensional variables, support structural equation modeling, and involve dimension reduction. They all use two-tier models where higher-order (latent or emergent) variables are inferred from lower-order (less problematic) variables.  They differ in their assumptions about the nature of the higher-order variables and the way they're estimated.

Choosing among these approaches is difficult because leading experts disagree about their relative merits and the circumstances under which each is appropriate. Our aim here is to advocate for measurement models in general and highlight some of the key issues involved. We will focus on the common factor model not because it is best, but because we understand it better than the others.

\subsection{The Common Factor Model}

The common factor model, like realism, posits that unobservable structures (like personality) \textit{cause} observable effects (like specific human behaviors) in our world.  

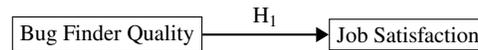
\begin{figure}
\centering
\begin{tikzpicture}
  \node[draw, rectangle] (BugFinderQuality) at (-2,0) {Bug Finder Quality};
  \node[draw, rectangle] (jobSatisfaction) at (2,0) {Job Satisfaction};
  
    \draw [thick,-{Triangle[fill=black]}] (BugFinderQuality.east) -- node [midway, above] {H\textsubscript{1}} (jobSatisfaction.west);
\end{tikzpicture}
\caption{Hypothesized Latent Structure}
\label{fig:tool_satisfaction}
\end{figure}

Imagine we are studying the job satisfaction of software testers. Suppose for simplicity that each tester uses just one of many available tools for identifying bugs. We hypothesize that the quality of the bug finding tool affects their job satisfaction (Figure~\ref{fig:tool_satisfaction}). The single arrow indicates the hypothesized causal relationship. 

\begin{figure}
\centering
\begin{tikzpicture}
  \node[draw, rectangle] (BFQ) {Bug Finder Quality};
  \node[draw, rectangle,right of=BFQ,xshift=4cm] (JS) {Job Satisfaction};
  \node[below left of=BFQ,yshift=-1cm] (BFQ2) {BFQ\textsubscript{2}};
  \node[left of=BFQ2] (BFQ1) {BFQ\textsubscript{1}};
  \node[right of=BFQ2] (BFQdots) {\dots};
  \node[right of=BFQdots] (BFQn) {BFQ\textsubscript{m}};
  \node[below left of=JS, yshift=-2cm] (JS2) {JS\textsubscript{2}};
  \node[left of=JS2] (JS1) {JS\textsubscript{1}};
  \node[right of=JS2] (JSdots) {\dots};
  \node[right of=JSdots] (JSn) {JS\textsubscript{n}};
  
  \draw [thick,-{Triangle[fill=black]}] (BFQ.east) -- node [midway, above] {H\textsubscript{1}} (JS.west);
  \draw [thick,-{Triangle[fill=black]}] (BFQ.south) -- (BFQ1.north);
  \draw [thick,-{Triangle[fill=black]}] (BFQ.south) -- (BFQ2.north);
  \draw [thick,-{Triangle[fill=black]}] (BFQ.south) -- (BFQn.north);
  \draw [thick,-{Triangle[fill=black]}] (JS.south) -- (JS1.north);
  \draw [thick,-{Triangle[fill=black]}] (JS.south) -- (JS2.north);
  \draw [thick,-{Triangle[fill=black]}] (JS.south) -- (JSn.north);
\end{tikzpicture}
\caption{Hypothesized Latent Structure with Reflective Indicators}
\label{fig:tool_satisfaction2}
\end{figure}
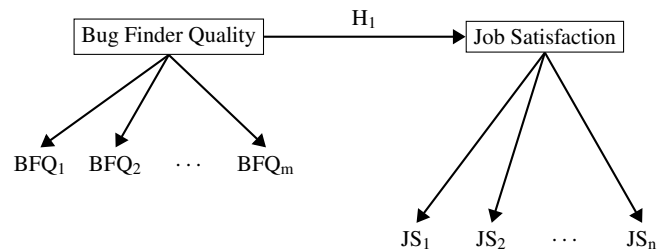

The technical phenomenon of tool quality, the psychological phenomenon of job satisfaction, and the causal structure linking them are \textit{all} latent; that is, their measurement is intrinsically problematic. To measure our two latent variables, then, we need to find some properties that are less problematic (i.e., less theory- and value-laden) to observe. In the common factor model, these less-problematic variables are called \textit{reflective indicators} because they reflect the latent variable of interest. We can then formulate a measurement model~\cite{Henseler2021} such as the one shown in Fig.~\ref{fig:tool_satisfaction2}.

There's a lot going on in this measurement model. You can see our hypothesis, \textit{H\textsubscript{1}}, that Bug Finder Quality causes Job Satisfaction. Bug Finder Quality (BFQ) is operationalized using $m$ reflective indicators labeled BFQ\textsubscript{1} \dots BFQ\textsubscript{m}. Job Satisfaction is operationalized using $n$ reflective indicators labeled JS\textsubscript{1} \dots JS\textsubscript{n}. (Naming the indicators like this helps keep your datasets organized later.) 

See how the arrows point \textit{from} the latent variable \textit{to} the reflective indicators? That's because the common factor model assumes that the latent variable that we can't see \textit{cause} changes in the indicators that we can see. The changes in the indicators \textit{reflect} changes in the underlying construct~\cite{Petter2007}.

We estimate the latent variable in terms of the \textit{shared variance}\footnote{A mathematically simple way to get the shared variance is to sum the indicators but more sophisticated approaches, such as confirmatory factor analysis, are typically used.} of its reflective indicators because the latent variable is the factor that all the indicators have in common---hence the name ``common factor model''.  

The usefulness of our measurement model therefore depends on (1) the quality of our reflective indicators and (2) the way we estimate our latent variables from them.

\begin{exercises}
    \textit{Select two or more constructs from the list of things you'd like to measure. Identify at least three reflective indicators for each. Draw a measurement model showing your constructs, indicators, and hypotheses}

     Students may need help identifying potential reflective indicators. Discussing their models in groups may help.  
\end{exercises}

\subsection{Good Reflective Indicators}

The best kind of reflective indicators are those comprising comprehensive instruments that have been used widely and validated repeatedly in lots of different contexts similar to the context at hand. Making and validating instruments is quite difficult, so using existing ones saves time and effort. Fortunately, good instruments exist for many psychological and psycho-social constructs.  For example, it's not hard to find a questionnaire instrument (AKA a \textit{scale}) for job satisfaction that has been used in dozens of studies and undergone both quantitative and qualitative evaluations of its psychometric properties. Unfortunately, good instruments do not exist for many technical constructs like the quality of a bug finding tool. 

In a good scale, the reflective indicators:
\begin{itemize}    
    \item measure different things that are all driven by the construct
    \item are highly but not perfectly correlated with each other (convergent validity);
    \item are not too correlated with indicators of other scales in the measurement model (discriminant validity); 
    \item cover all aspects or dimensions of the phenomenon of interest (content validity).
\end{itemize}

Good questionnaire scales have additional properties. For example, including both direct (e.g. ``My work is interesting'') and reversed items (e.g. ``I feel bad about my job'') improves construct validity and mitigates response bias. While many software metrics are also reversed (e.g. LCOM metrics measure \textit{lack of }cohesion), it probably doesn't matter because the ostensible benefits of including reversed items are all about the psychological effect they have on a person filling out a questionnaire. Regardless, inverting the values of reversed indicators at the beginning of your analysis helps avoid confusion. 

Technical metrics can be so highly correlated that they cause multicollinearity problems and some statistical analyses will fail. Multicollinearity can be addressed by dropping one or more indicators. If two indicators are nearly perfectly correlated, you don't lose any meaningful information by dropping one of them. 

Beyond that, the meanings of ``highly correlated'' and ``not too correlated'' are relative to context. One approach is to generate a correlation matrix for all of your indicators. Then, divide them into two groups: correlations between items in the same scale (A), and correlations among items in different scales (B). The \textit{smallest} member of A should be \textit{larger} than the \textit{largest} member of B. When doing this comparison, either take the absolute value of all the correlations or reverse any reversed indicators before you start.  

However, you can't just ask the same question or measure the same thing \textit{n} different ways. For example, you would not have a questionnaire with items like ``I am happy with my job''; ``I am pleased with my job''; ``I like my job''; ``My job is great,'' etc. These semantically identical items will create the illusion of validity.  You need semantically diverse indicators, the answers to which will all be driven by the same construct. For example, some of the items in MacDonald and Maclntyre~\cite{macdonald1997generic} scale are ``I feel secure about my job'', ``My wages are good'' and ``I get along with my supervisors.'' These are semantically different questions.

Similarly, to measure the size of a software system, we don't count lines of code, non-comment lines of code, logical lines of code, lines of code including third party libraries; lines of code excluding third party libraries, and so on. These metrics are too similar. We need to count \textit{different} things that are all driven by the size of the system like number of methods, number of fields, number of classes, one lines-of-code variant, etc. 

These items should cover all aspects of the latent variable. With unidimensional variables, this just means asking lots of semantically diverse questions or selecting diverse technical metrics. Multidimensional variables are a little trickier. 

\subsection{Multidimensional Variables}

When we operationalize a construct using several reflective indicators, it's called a \textit{reflective measurement model}. Modeling something like software quality reflectively is problematic because it has several dimensions that may be weakly correlated (e.g. usability and carbon footprint) or inversely correlated (e.g. precision and recall, effectiveness and efficiency). 

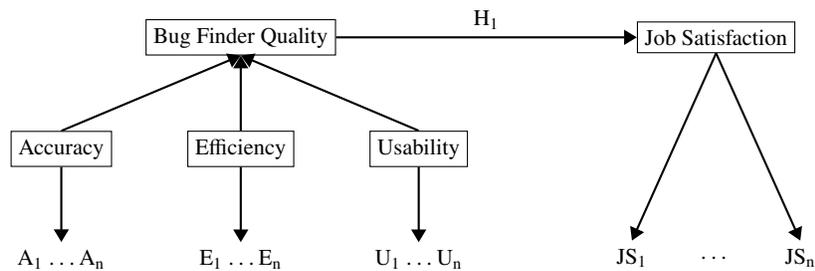
\begin{figure}
\centering
\begin{tikzpicture}
  \node[draw, rectangle] (BFQ) {Bug Finder Quality};
  \node[draw, rectangle, right = 4cm of BFQ] (JS) {Job Satisfaction};
  \node[draw, rectangle, below = of BFQ] (Efficiency) {Efficiency};
  \node[draw, rectangle, left = of Efficiency] (Accuracy) {Accuracy};
  \node[draw, rectangle, right = of Efficiency] (Usability) {Usability};
  \node[below = of Accuracy] (A1n) {A\textsubscript{1} \dots A\textsubscript{n}};
  \node[below = of Efficiency] (E1n) {E\textsubscript{1} \dots E\textsubscript{n}};
  \node[below = of Usability] (U1n) {U\textsubscript{1} \dots U\textsubscript{n}};
  \node[below = 2.6cm of JS] (JSdots) {\dots};
  \node[left = 0.5cm of JSdots] (JS1) {JS\textsubscript{1}};
  \node[right = 0.5cm of JSdots] (JSn) {JS\textsubscript{n}};
  
  \draw [thick,-{Triangle[fill=black]}] (BFQ.east) -- node [midway, above] {H\textsubscript{1}} (JS.west);

  \draw [thick,-{Triangle[fill=black]}] (Accuracy.north) -- (BFQ.south);
  \draw [thick,-{Triangle[fill=black]}] (Efficiency.north) -- (BFQ.south);
  \draw [thick,-{Triangle[fill=black]}] (Usability.north) -- (BFQ.south);
  
  \draw [thick,-{Triangle[fill=black]}] (Accuracy.south) -- (A1n.north);
  \draw [thick,-{Triangle[fill=black]}] (Efficiency.south) -- (E1n.north);
  \draw [thick,-{Triangle[fill=black]}] (Usability.south) -- (U1n.north);
  \draw [thick,-{Triangle[fill=black]}] (JS.south) -- (JS1.north);
  \draw [thick,-{Triangle[fill=black]}] (JS.south) -- (JSn.north);
\end{tikzpicture}
\caption{Hypothesized Latent Structure with Multidimensional Variable}
\label{fig:tool_satisfaction3}
\end{figure}

Suppose we have three latent variables: Accuracy, Efficiency, and Usability (of the Bug Finder tool), and a \textit{second order} latent variable, Bug Finder Quality, as shown in Figure~\ref{fig:tool_satisfaction3}.

Further suppose we have separate, multi-item scales (or multi-metric instruments) for each of Accuracy, Efficiency, and Usability. These may include subjective measures (e.g., of perceived efficiency), objective measures (e.g. benchmark results), or both.

If the reflective indicators for Accuracy, Efficiency and Usability are all highly intercorrelated, we don't need to model them as separate dimensions~\cite{Fassott2015}. We just combine all three subscales, and go back to the model shown in Fig.~\ref{fig:tool_satisfaction2}. However, if Accuracy, Efficiency, and Usability aren't highly correlated, or are inversely correlated, we need something else. 

And here is where it gets messy. This situation is difficult to model in covariance-based SEM, which is probably why formative models are underused (i.e. formative constructs are often modeled as reflective)~\cite{Fassott2015}. Some people use ``Mode B'' in PLS to model formative relationships, but what PLS is actually doing, mathematically, may or may not be a sensible way to model the formative relationship in question. A host of other approaches to modeling composite and emergent variables have been proposed (e.g. forged concepts~\cite{Liu2022Forged}) but there is no consensus about when to use what. 

Here's what we can say with confidence. A causal-formative measurement model posits that the dimensions (e.g. efficiency) \textit{cause} the higher-order construct (e.g. Bug Finder Quality)~\cite{coltman2008}. Thus, the ideas of convergent and discriminant validity are less helpful in formative models. Whereas reflective measurement models consider measurement error at the indicator level, formative measurement models consider measurement errors at the construct level. When the only multidimensional variable in an experiment is the dependent variable, and it has a small number of dimensions (e.g. $d\leq5$) it may be preferable to run $d$ separate structural equation models---one for each dimension.  

Further mathematical details of the different approaches to estimating second-order latent variables are beyond the scope of both this chapter and introductory research methods courses. Here, we just want to emphasize that (1) including multidimensional constructs significantly increases analytical complexity; and (2) students should not invent their own half-baked math for estimating composite constructs. 

\begin{exercises}
    \textit{Look over your list of things you'd like to measure. Are any of them multidimensional? What are their dimensions?}
\end{exercises}

\subsection{Lessons Learned}

Most software engineering students are comfortable enough with mathematics and algorithms to quickly the idea of operationalizing a latent variable as the shared variance among several reflective indicators. They quickly grok that it's like a weighted average. However, many students struggle to apply this idea to their work, especially when it's not the norm in their subfield. For example, many software engineering researchers build tools to identify bugs automatically, and evaluate their effectiveness using a measure of recall. They don't even bother with precision because they reason that the user can manually disregard false positives. Recall isn't latent and the case for multiple measures is murky. 

We can address this by focusing on \textit{their concerns} rather than a principled argument about rigor. Researchers worry that their studies won't produce any significant results and will be harder to publish. If the researcher backs up the conceptual hierarchy and chooses a more multifaceted (and latent) view of success, it creates more opportunities for significant results\footnote{Even when correcting $\alpha$ for multiple comparisons, the more success dimensions we evaluate, the more likely we will find one on which new tool excels}. Perhaps their bug fixing tool will underperform existing tools on recall, but is more efficient, has higher precision, and is easier to incorporate into commercial build systems. Showing that the new system outperformed previous systems \textit{on some dimension} increases the chances of publishing the paper.

Selecting papers based on whether statistical significance was achieved, is, of course, totally unscientific. Regardless, more comprehensive success measures are good for science. For example, making software more environmentally sustainable is much more difficult if sustainability is not considered as part of success. 

Despite usually being good with math, students universally struggle with the idea of evaluating convergent and discriminant validity by comparing the smallest intra-scale correlation to the largest inter-scale correlation. Instructors will need to demonstrate, and students should try this themselves in an assignment or lab setting.

\section{Instrumentation and Scaling}
\label{sec:intrumentation}

\begin{learningObjectives}
    \begin{enumerate}
        \item Explain how to create a multi-metric scale to measure a property of a software system 
        \item Use factor analysis to assess and improve the convergent and discriminant validity of a unidimensional, multi-item scale or multi-item instrument 
    \end{enumerate}
\end{learningObjectives}

\subsection{Instrumentation for Psychometric Properties}

One of the most popular (and often misunderstood) methods of creating a questionnaire instrument to measure a psychological attribute was proposed by Likert~\cite{briggs2021historical}. A \textit{Likert scale} is \textbf{not} asking a respondent to agree or disagree with a statement on a five-point scale. Instead, a Likert scale is a multi-item summative scale designed using Likert's rigorous scaling method. 

Many detailed accounts of Likert's scaling method are available (e.g.~\cite{trochim2016Research}) so we won't repeat the process here. Briefly, though, Likert scaling involves clearly defining the construct of interest, generating 80--100 possible items, and multiple rounds of pilot testing and statistical analysis to select the best subset of items. Further rounds of qualitative analysis to assess the psychometric properties of the scale are recommended. This process can require 30 or more human participants. Once the resulting scale has been used in studies and shown to have predictive validity (i.e. predict or be predicted by what we theorized), we are even more confident in our operationalization\footnote{``Operationalize'' is a common term in the literature on construct validity. It refers to how we measure the construct, including our instruments and statistical approach.} of the construct.

Alternative psychometric scaling approaches including Thurstone, Guttman, and semantic differential are equally complicated. Simply making up some questions is not acceptable. 

\subsection{Instrumentation for Software Properties} \label{ssec:Instrumentation}

This section describes one good way of developing an instrument to measure a property of a software system. This is not the only way; rather, it exemplifies the level of rigor that should pervade instrumentation. Our recommended approach looks labor intensive because it is. However, when researchers apply such rigorous instrumentation approaches, they typically get to write an instrument development paper (which becomes part of a graduate student's thesis), and have more confidence in their subsequent findings.  

\subsubsection{Define the Constructs of Interest}

First we need a clear definition of the construct we want to measure. Common words with multiple meanings like ``complexity'' or ``understandability'' are not sufficient. For example, defining maintainability as ``the ease with which a software system can be modified, repaired, or enhanced over time'' would be problematic because it sets up maintainability as a mutual property of a system and a person (the maintainer). You need to think deeply about what \textit{exactly} you're trying to measure. Ask yourself not only `what am I trying to measure?' but also `from whose perspective?' and `in what context?'. What are the different kinds or dimensions of the property you want to measure? What is it's range? Try to define it using common everyday language. If you must use jargon, the jargon must also be clearly defined in common, everyday language or technical terms that have standard or widely-agreed meanings. 

\begin{exercises}
    \textit{Choose one of the constructs you'd like to measure and define it as precisely as possible. Discuss your definition in small groups. Look for ambiguity in each others' definitions.}
\end{exercises}

\subsubsection{Select Related Constructs}

Assessing construct validity works better when our construct is situated withing a small group of similar constructs. This helps us  ensure that our construct differs significantly from related constructs. For example, when measuring the complexity of a software system, we'd want to ensure that complexity wasn't conflated with system size, coupling, or cohesion. If one of our coupling metrics is correlated more with some cohesion metrics than with our other coupling metrics, we'd get suspicious. Therefore, we should select a few (no more than five) related constructs. If we can think of more than five, select the most closely related. 

It's ok if we believe these construct are correlated or causally related. Correlation between constructs doesn't mean that they measure the same thing. 

It's not ok if two constructs are essentially the exact same thing. For example, we would not include ``trust in organization'' and ``perceived risk of dealing with organization'' because these are arguably semantically equivalent. 

\begin{exercises}
    \textit{Brainstorm a small set of related constructs.}
\end{exercises}

\subsubsection{Select, Vary, or Create Metrics}

Software engineering researchers have proposed many metrics that propose to measure important properties of software systems. For example, productivity metrics include lines of code changed, function points implemented, number issue tickets solved, number of commits made, and number of files changed. Using a well-validated, multi-metric instrument for one of the related constructs will save time and help your work build on the cumulative body of knowledge. 

For each construct that doesn't have well-validated, multi-metric instrument available, generate a long list of possible metrics. The exact number depends on how complex the construct is, but aim for \textit{at least} 20. We can generate metrics in at least three ways:
\begin{enumerate}
    \item Find metrics that have already been proposed for this metric.
    \item Create new metrics that seem like plausible indicators of the construct.
    \item Create variations.
\end{enumerate}

For example, suppose we want to measure the size of an object-oriented code base. First we look at metrics calculated by existing tools or proposed by prior research: non-comment lines of code, numbers of classes, methods, fields, etc. If there are many such metrics and they seem to cover the entire range of the construct, perhaps that's sufficient. However, brainstorming novel metrics is a useful exercise even if it seems like we have enough. For size, we might imagine things like number of unit tests, lines of documentation, or number of user stories implemented. We continue creating new, diverse metrics until we have good content validity; that is, we've covered all areas or dimensions of the construct.

Each metric we find or create should be scrutinized for face validity---does this make an sense? Using app store star ratings as a code quality metric doesn't make sense ``on it's face'', because most of the people leaving those ratings never see the code.  

Once we are confident in our face and content validity, we can generate variations. For example, when computing non-comment lines of code in Java we could have variations that include or exclude private methods, inner classes, enums, the standard library, or third-party libraries. That's $2^5=32$ variations. \textbf{DO NOT} create 32 variations for each metric. You'll increase the complexity of subsequent steps, and eventually have to drop most variations due to multicollinearity problems anyway. DO include some variations that make sense theoretically. For example, when measuring coupling including third-party libraries makes more sense than including the standard library~\cite{TEMPERO2018Coupling}.

Generating variations will not substantially improve content validity. Indeed, only one variation is likely to survive the subsequent weeding process. Including variations is more about improving the quality of the resulting instrument by ensuring the best variation of each metric is used.  

At this stage you should include a metric even if you worry that it conflates two different constructs, or you're not sure which of your constructs it measures. Those issues will be resolved below. 

\subsubsection{Collect Data}

We need to collect approximately ten observations per metric for a good factor analysis~\cite{Hair2013Multivariate}. For example, if we have 152 class-level Java code metrics, we need about 1520 classes. We don't need a representative sample of all the world's Java classes---purposive sampling is fine at this stage~\cite{baltes2022sampling}. It will be easier if all of the classes come from the same project. Not just any project will do though. For example, if you have metrics that include enums, you need a project that has some enums. Once you've collected your sample of source code, calculate all the metrics.

\subsubsection{Exploratory Factor Analysis}

Factor analysis is a technique used to extract ``factors'' or ``constructs'' or ``latent variables'' from a set of ``variables'' or ``reflective indicators'' to reveal the underlying latent structure. Exploratory factor analysis (EFA) analyses the correlation between variables in a dataset to estimate which factor they would belong to. The higher the correlation between variables, the more likely they are to be measuring the same underlying factor. There are two kinds of factor analysis: exploratory and a confirmatory. As the name suggests, an exploratory factor analysis (EFA) helps us explore which indicators reflect which factors, while confirmatory factor analysis (CFA) provides ``confirmation'' that the factors do represent the correct variables~\cite{Hair2013Multivariate}.

We can use exploratory factor analysis (EFA) to explore the underlying factor structure of the data; that is, the degree to which metrics converge with other metrics for the same construct and diverge from metrics associated with other constructs. To reveal factors through EFA, we need many reflective indicators that ostensibly measure each construct and capture all aspects of each construct. That's why we need lots of diverse metrics, not just a few metrics or a few variations on a single metric.  

Guidelines for conducting an EFA are available elsewhere~\cite{Hair2013Multivariate}. We include a worked example (Appendix B) and provide a sample dataset and scripts that can be used for demonstrations or lab assignments (see \nameref{sec:Supplements} on page \pageref{sec:Supplements}). Needless to say, we do not simply compute and EFA and call it a day. We have to iteratively refine the model until convergent and discriminant validity are high. 

All of our example data is numeric. A factor analysis only accepts numerical data in its analysis, so we'll have to recode any non-numerical data to use it. We only consider uni-dimensional measures at the class level. Multi-dimensional measures require more sophisticated approaches (as discussed above); measures at different levels may require a multi-level modeling approach. 

This exploratory part of the instrument development process includes many subjective decisions. Many of these decisions have no firm theoretical basis---we just have to make a reasonable decision and move on. Don't worry about p-hacking or over-fitting at this stage. 

Once we're finished tweaking the EFA model, we should re-assess content validity. That is, we should ensure that the remaining metrics cover all aspects of the constructs at hand. If so, the metrics included in the final EFA model are the only ones we need going forward. If not, we need to add some new metrics and try again.

\subsubsection{Confirmatory Factor Analysis}

At this point, we have a tentative measurement model, but it was devised subjectively, probably using data from a single project. Many of our choices cannot be justified beyond seeming reasonable at the time. So we want to confirm that our model holds up on a different and larger dataset. 

Now sampling matters. We want a large, sample of all the world's code that we can argue is representative (see~\cite{baltes2022sampling}). (Note: we don't want snapshots of the same project across many different times---that may cause problems.) But we don't have a \textit{sampling frame}: an index of all the world's code to sample from. So we have some options:
\begin{itemize}
    \item Use a standard code corpus like Qualitas~\cite{tempero2010qualitas} or PyTorrent~\cite{bahrami2021pytorrent}.\footnote{We have not used PyTorrent but it looks promising.} The advantage of a corpus is that our work is easier to replicate and compare to other studies using the same corpus. The disadvantage is that corpus is probably biased toward higher-quality code, while our measures should work (and be validated) on both good and bad code.
    \item Get codebase from a large company (or if the company can't share the code, get them company to run the analysis for you and send you the results). The problem here is every company is unique. We cannot generalize from Microsoft's code to Apple's or vice versa. 
    \item Sample randomly from Github. This has the strongest argument to representativeness because of the diversity of projects hosted on Github at the time of writing. However, there is a possibility that open source code systematically differs from closed-source code and that our study will be harder to replicate.
    \item Get two samples: one more diverse open-source sample and one less diverse closed-source sample from a partner organization. Do the CFA twice and compare the results. This is the most rigorous approach we can think of, but it's more laborious. 
\end{itemize}

\begin{exercises}
    \textit{Look at the list of things you want to measure. What kinds of organizations might care about measuring these things? Might any of them care enough to work with you? Do you know anyone at those organizations, or who could introduce you to people in those organizations?}
\end{exercises}

Next, we calculate all of the metrics in our final EFA model and run confirmatory factor analysis (CFA). For CFA, we must specify not only the number of factors, but also which metrics correspond to which factors. Comprehensive guidelines for conducting CFA are available elsewhere~\cite{harrington2009confirmatory}, so here we will just include a few notes:
\begin{itemize}
    \item Like EFA, CFA produces factor loadings that indicate how strongly each metric relates to its factor.
    \item Unlike EFA, CFA only produces loadings for the construct a metric is assigned to. 
    \item EFA loadings range from -1 to 1. CFA loadings can be greater than 1, and that a good thing!
    \item CFA produces estimates of each constructs called ``factor scores'' that we can use for causal analysis. 
\end{itemize}

CFA produces a new measurement model. It has the same metrics as the EFA model, but a different formula for estimating the constructs from the metrics. These formulas collectively make up our measurement model. Now we can use these formulas to estimate our constructs of interest without repeating the CFA (or any previous steps in the instrument development process). 

During our EFA analysis, we can fiddle with the model as much as we want. But then we move from an exploratory phase to a confirmatory phase. This means we only run the CFA once, and then we write our instrument development paper and report our results. If some of our metrics don't load well, that means our measurement model is only ``partially supported.'' In the discussion section of our paper, we can make recommendations like ``consider dropping metric $m$'' but we don't iterate at this stage because then it wouldn't be confirmatory testing anymore. 

Later, when conduct a study \textit{using} our instrument in causal analysis, we might find one or more metrics load poorly. \textit{Then,} we can drop problematic metrics. Dropping problematic metrics is pretty common, which is one reason why it's useful to have extra metrics in our measurement model.

\subsubsection{Additional steps}

By the end of the CFA process, we should be very confident in our measurement model. However, our confidence increases if our measure demonstrates measurement invariance (Section~\ref{ssec:MeasurementInvariance}) and predictive validity. For example, if we had instruments for the size and complexity of a software system, and we found that systems became \textit{less} complex as they grew, we would suspect that something was wrong with our measures. 

However, we would argue that the time to write an instrument development paper is at the end of the CFA. Measurement invariance and predictive validity, in contrast, should be assessed when we use our new instruments to do some causal analysis. Therefore, further exploration of these topics is beyond the scope of this chapter. 

\subsection{Lessons Learned}

Instrument development is easier to understand if you do it rather than just talking about it. For psychometric instruments, leading students through an elaborate simulation in which they actually create a questionnaire scale is an excellent way of conveying the process. This simulation must be performed over several classes because the instructor needs time for additional tasks, such as deduplicating the items. 

The most difficult part is rating the favorability of each item toward the concept. For example, the item ``My work makes me sad'' has low favorability toward the concept ``job satisfaction.'' Students who are not paying attention to the exercise sometimes rate it low because ``My work makes me sad'' sounds bad, or because \textit{their} work doesn't make \textit{them} sad. The instructor should go through some examples like this one to ensure the students understand the task. 

As should be clear from the preceding guidelines, generating an instrument to measure a technical property of a software system is a lot to simulate in a single class. Instead, we recommend giving an overview of the process, followed by lab assignment corresponding to discrete steps, and the online supplement includes materials that can be used for an EFA demonstration or lab assignment. 

\section{Discussion and Conclusion} 
\label{sec:conclusion}
\subsection{Summary}

This chapter explores a broad spectrum of measurement issues in the context of software engineering research. Its core message is that software engineering research is being held back by a lack of attention to measurement. Notwithstanding some recent, excellent papers about measurement (e.g.~\cite{graziotin2021psychometrics,sjoberg2022construct,russo2021pls}), much SE research is not trustworthy because it uses unreliable, unvalidited measures. 

We began, in Section~\ref{sec:causalWeb}, by discussing the importance of selecting appropriate variables from the causal web surrounding one's phenomenon of interest. We argued that choosing variables that are too proximate reduces impact, while choosing variables too remote impedes research. Once we've selected variables, we must find ways of measuring them \textit{reliably}. In Section~\ref{sec:representational}, we argued for conceptualizing reliability in terms of true score theory, quantitatively assessing reliability and measurement invariance, and improving reliability through piloting, triangulation, and double-entering data. 

Reliability, is necessary but insufficient for validity. That is, you can consistently measure the wrong thing. Section~\ref{sec:realism} therefore introduced the realist view of \textit{validity}---the degree to which a given instrument measures what it is supposed to measure. We argued that measurement is intrinsically problematic because observable changes in the world are often driven by unobservable structures and processes. This necessitates statistical measurement models, as explained in Section~\ref{sec:modelbased}. We focused on reflective, common-factor models, only touching upon the more difficult cases of multidimensional and causal-formative measurement models. The main takeaway of this section is that practically all technical properties of software systems and psycho-social properties of software developers(/teams/organizations) require sophisticated statistical measurement models and should not be operationalized as singular ``proxy'' variables. Section~\ref{sec:intrumentation} therefore provides recommendations for building statistical measurement models for psycho-social and technical properties. Good instrumentation is arduous and, we argue, should constitute sufficient contribution for a full-length technical journal article. 

\subsection{Lessons Learned: Assessment Strategies}

Assessing students' understanding of the concepts described above can be challenging. In some universities, research methods are taught in workshops, seminars or lab meetings with little assessment or grading. Others have conventional courses with projects or exam-based assessment. At the undergraduate level, exams can be used to assess superficial knowledge of measurement (e.g. ``give examples of metrics''; ``define systemic error''; ``explain the importance of measurement invariance''). At the graduate level, however, exams cannot assess the ability to conceptualize latent structures, generate an appropriate measurement model, and quantitatively analyze construct validity.

Many graduate research methods courses end with writing a research proposal. In principle, working out the measurement strategy is a core part of a research proposal, so this kind of assignment should work. In practice, most students struggle to construct a research proposal in sufficient detail. For example, when proposing an experiment with human participants, students tend to produce hand-waving descriptions of procedures and materials rather than usable study protocols and task materials that participants could actually receive, understand, and complete. Students struggle to formulate appropriate statistical procedures including contingencies (e.g. ``what will you do if the residuals are not normally distributed?''). 

We therefore have the following suggestions for alternative assessments. We would not expect an introductory research methods course to include all of these assessments; these are just possibilities. 
\begin{itemize}
    \item Write a detailed critique of a given instrument development paper.  
    \item Create a rubric for evaluating instrument development papers similar to the SIGSOFT Empirical Standards~\cite{ralph2021empirical}
    \item A series of labs and corresponding lab assignments in which students complete the steps in Section~\ref{ssec:Instrumentation}. 
    \item \textit{(graduate level only)} Find a research paper with a comprehensive replication package including a complete data set and a statistical measurement model (e.g.~\cite{ralph2020pandemic}) and replicate the analysis. Suggest (or better yet, implement) specific improvements to the measurement model. 
\end{itemize}

\subsection{Conclusion}

In conclusion, we stress the need for software engineering researchers to pay more attention to measurement, quantitatively assess reliability and validity, embrace statistical measurement models, and develop more high-quality scales and instruments. We call on reviewers not only to raise their expectations around measurement (e.g. when reviewing papers, proposals, and grant applications) but also to give more credit when measurement is done well, and to recognize that creating and validating an instrument is a study on its own. Lastly, we call on educators who teach software engineering research to provide more explicit instruction around measurement issues, and we hope this chapter is a strong resource for helping in this regard. 

\section*{Supplementary Materials} \label{sec:Supplements}

Supplementary materials for the reliability and exploratory factor analysis---including a dataset, sample scripts, sample results, and definitions of code quality metrics---can be found at \url{https://doi.org/10.5281/zenodo.11544897}. 

\ethics{Competing Interests}{The authors have no conflicts of interest to declare that are relevant to the content of this chapter.}

\bibliographystyle{spmpsci}
\bibliography{bibliography}

\section{Appendix A: Reliability Analysis}
\label{sec:relAnalysis}

This Appendix provides an example of analyzing reliability on software metrics computed by different tools. The metrics were calculated from the source code of \textit{Apache Maven}.\footnote{https://github.com/apache/maven} You can find the data and scripts in the online supplement to this book (see \nameref{sec:Supplements} on page \pageref{sec:Supplements}). 

The purpose of this analysis is to investigate the extent to which metrics calculated by different tools provide consistent measurements.

\subsection{Data Preparation}
\label{ssec:prepData}

The dataset includes various metrics such as size, cohesion, inheritance, and coupling metrics, which are continuous. The following steps were undertaken to prepare the data for reliability analysis:
\begin{enumerate}
    \item Read the data from the excel file containing the metrics.
    \begin{verbatim}
    library(readxl)
    data <- read_excel("efaReadyMC.xlsx")
    \end{verbatim}
    \item Select the relevant metrics for analysis. Here we use Lines Of Code (LOC) Metrics computed by three different tools---Designite\footnote{\url{https://www.designite-tools.com/}}, JHawk\footnote{\url{http://www.virtualmachinery.com/jhdownload.htm}}, and Understand\footnote{\url{https://scitools.com/}}. The corresponding columns in the dataset are Size.LOC.Designite, Size.LOC.JHawk, Size.LOC.Understand.
    \begin{verbatim}
    rel1_data <- 
        select(rel_data, 'Size.LOC.Designite', 
        'Size.LOC.JHawk', 'Size.LOC.Understand')
    \end{verbatim}
    \end{enumerate}

\subsection{Calculate a Measure of Reliability}
\label{ssec:omega}

Since lines-of-code is ratio-level data, we'll use a measure of reliability rather than agreement (see Section~\ref{ssec:IRR}). For this example, we will use Cronbach's alpha because it is simpler to calculate an interpret. Note, however, that if we were looking at reliability after doing factor analysis or a similar technique, more accurate measures of reliability such as McDonald's omega and Composite Reliability are available. 

We will calculate Cronbach's alpha using the \texttt{psych} package\footnote{https://personality-project.org/r/psych/} in R as follows.
\begin{enumerate}
  \item Convert the selected data into a dataframe.
    \begin{verbatim}
    rel1_data <- as.data.frame(t(rel1_data))
    \end{verbatim}
   
    \item Calculate alpha
    \begin{verbatim}
    library(psych)
    alphaResult <- alpha(rel1_data)
    \end{verbatim}
 \end{enumerate}

\subsection{Results and Interpretation}

Cronbach's alpha values range from -1 to 1, with higher values indicating greater reliability and internal consistency among the measured items. Generally, $\alpha>0.7$ is considered acceptable, while $\alpha>0.9$ is considered excellent. Our result, $\alpha=0.97$ indicates excellent reliability. This suggests that the three tested tools are measuring basically, if not exactly, the same thing. However, excellent reliability doesn't mean that the studied metrics reflect the target underlying construct (e.g. class size). To determine that, we need a different kind of analysis (next). 

\addcontentsline{toc}{section}{Appendix}

\section{Appendix B: Exploratory Factor Analysis} \label{sec:AppendixEFA}

This Appendix provides an example of an Exploratory Factor Analysis, following established guidelines~\cite{Hair2013Multivariate} and using selected metrics calculated from the source code of \textit{Apache Maven}\footnote{https://github.com/apache/maven}. You can find the data and scripts in the online supplement to this book (see \nameref{sec:Supplements} on page \pageref{sec:Supplements}). 

\subsection{Objective of Factor Analysis}

The objective of our exploratory factor analysis is to assess the convergent and discriminant validity of common, object-oriented, class level, software code quality metrics calculated on the source code of Apache Maven where convergent validity refers to how similar the measure is with other measures it should be theoretically similar to and discriminant validity refers to how different the measure is with other measures it should theoretically be different to~\cite{trochim2016Research}.

\subsection{Design the Factor Analysis}

The dataset we will use contains measurements from 22 metrics---five size metrics, four cohesion metrics, four sup-inheritance metrics, four sub-inheritance metrics, two in-coupling metrics, and three out-coupling metrics. Our dataset includes measurements from approximately 1000 classes, well over the 10 observations per variable threshold. We aim to classify these metrics into six factors:
\begin{enumerate}
\item Cohesion: degree to which elements of a class belong together.
\item In-Coupling: degree to which a class is used by other classes. 
\item Out-Coupling: degree to which a class depends on other classes.
\item Size: how big the class is.
\item Sub-Inheritance: the degree to which a class has subclasses in an inheritance hierarchy.
\item Sup-Inheritance: the degree to which a class has superclasses in an inhreitance hierarchy. hierarchical properties related to super classes. 

\end{enumerate}

\subsection{Check Assumptions of Factor Analysis}

The assumptions of a factor analysis, and how we justify or test them, are as follows.

\begin{itemize}
\item Factor analysis should only be used when we theorize that a latent factor structure exists. In this case, we theorize that specific factors (size, coupling, etc.) are latent and do drive changes in metrics. 

\item Homogeneous sample of measurements. In other words metrics are calculated on the same sample of classes.

\item Multicollinearity. If none of the variables are correlated we cannot perform factor analysis; however, if two or more variables are perfectly or near perfectly correlated, it will cause a ``non-positive definite'' matrix, which will prevent the factor analysis from completing. We can assess multicollinearity in three ways: 
    \begin{enumerate}
    \item Visually inspecting a correlation matrix. In this case, we can see many correlations $>0.3$, which indicates that a factor analysis is possible~\cite{Hair2013Multivariate}. 
    \item The Kaiser–Meyer–Olkin (KMO) test. The KMO test tells us how correlated the variables in a dataset are. A minimum KMO value of 0.5 is acceptable and a value above 0.7 is recommended for a good factor analysis~\cite{Kaiser1974Little}. Our $KMO = 0.71$ is considered ``middling'' and appropriate for factor analysis~\cite{Kaiser1974Little}. The KMO values for each individual metric are also greater than the acceptable minimum of 0.5~\cite{Kaiser1974Little} (Fig. \ref{fig:kmo}).
    
\begin{figure}
\centering
\centerline{\includegraphics[width=\linewidth]{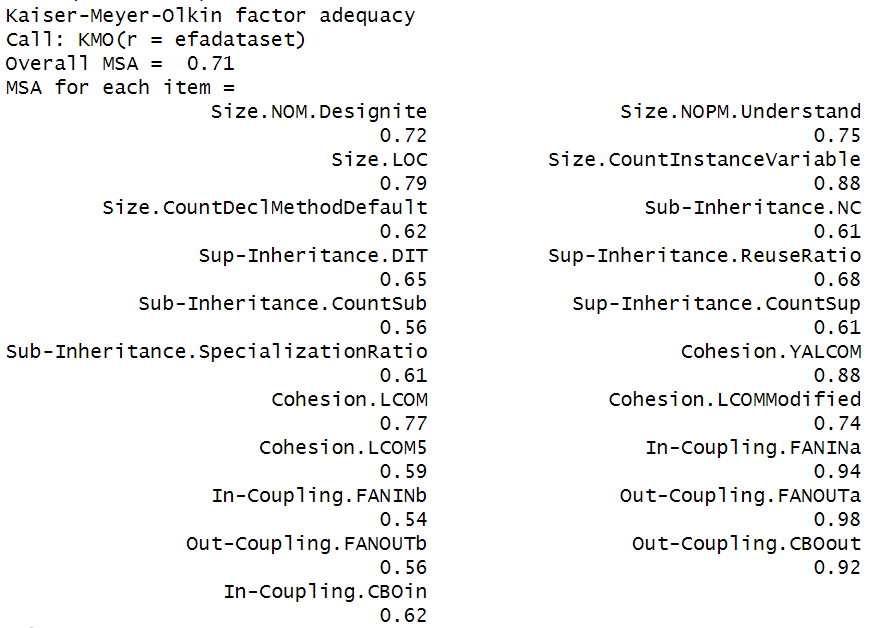}}
\caption{Kaiser–Meyer–Olkin (KMO) Test}
\label{fig:kmo}
\end{figure}

\item Bartlett's test of Sphericity - Bartlett's test of sphericity analyzes the correlations between variables to see if they are large enough to perform a factor analysis~\cite{Field2017Discovering}. It was found that Bartlett’s test of sphericity is significant ($p<0.001$) and we can proceed with the factor analysis (Fig. \ref{fig:bartlett}).

\begin{figure}
\centering
\centerline{\includegraphics[width=0.6\linewidth]{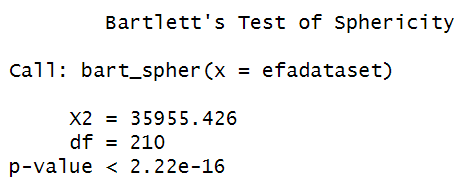}}
\caption{Bartlett's Test of Sphericity}
\label{fig:bartlett}
\end{figure}

\end{enumerate}
\end{itemize}

\subsection{Derive Factors and Assess Fit}

Researchers disagree on the best method of determining the number of factors to extract. We recommend using several methods to inform the decision~\cite{Costello2005Best}.

\begin{figure}
\centering
\includegraphics[width=0.9\linewidth]{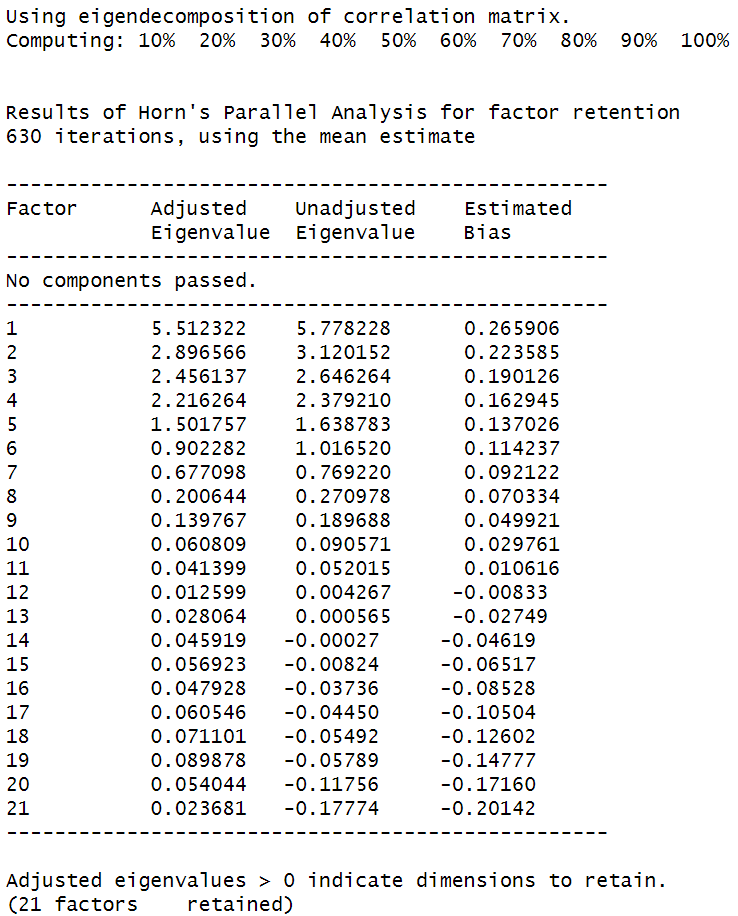}
\caption{Parallel Analysis}
\label{fig:parallelAnalysis}
\end{figure}

\textit{Parallel analysis} is a technique to estimate factors by calculating eigenvalues of random, uncorrelated data with eigenvalues of the actual data. The number of factors to retain is the count of eigenvalues greater than zero~\cite{Horn1965Rationale}. In our data, 21 factors are retained---21 eigenvalues are greater than zero (Fig. \ref{fig:parallelAnalysis}). Parallel analysis is known to overestimate the number of factors extracted when the dataset is very large~\cite{Costello2005Best} like ours.

\begin{figure}
\centering
\centerline{\includegraphics[width=\linewidth]{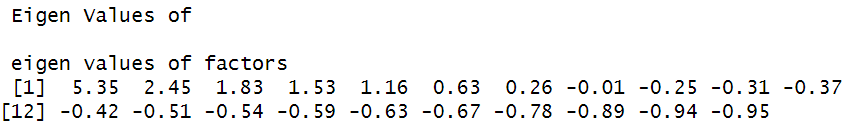}}
\caption{Kaiser Criterion}
\label{fig:kaiserCri}
\end{figure}

Alternatively, we retain a number of factors equal to the number of eigenvalues greater than one (the \textit{Kaiser Criterion}~\cite{Kaiser1960Application}). This method also loses effectiveness as the size of the dataset increases~\cite{Velicer2000Construct}, but not so much. We found five eigenvalues greater than one (Fig. \ref{fig:kaiserCri}), suggesting that we retain five factors.

\begin{figure}
\centering
\centerline{\includegraphics[width=\linewidth]{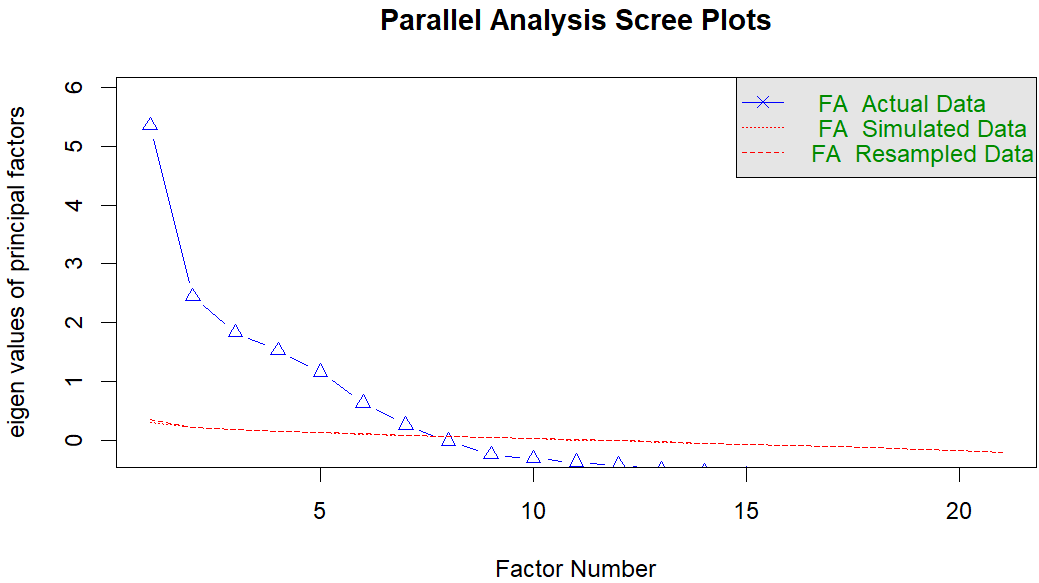}}
\caption{Scree Plot}
\label{fig:screePlot}
\end{figure}

Alternatively, we can estimate number of factors to retain by counting the eigenvalues before the bend in a \textit{scree plot}~\cite{Cattell1966Scree}. This technique is a little tricky and requires expertise if the plot is complicated~\cite{Costello2005Best}. Fig. \ref{fig:screePlot} has multiple bends---at 2, 4, and 7 eigenvalues---which suggests retaining anywhere between 2 and 7 factors.

In the \textit{theory} approach, we retain the number of factors that we theorize exist; in this case, six: size, cohesion, sub-inheritance, sup-inheritance, in-coupling, and out-coupling.

From the above discussion we have the following findings:
\begin{enumerate}
    \item Parallel analysis suggests 21 factors.
    \item The Kaiser criteria suggests 5 factors.
    \item The scree plot suggests factors between 2--7.
    \item Theory suggests 6 factors.
\end{enumerate}

Based on this, we tentatively retain seven factors as shown in Table \ref{tab:efa7f}. 
Table \ref{tab:efa7f} shows the variables and their corresponding factors loadings on each of the seven factors. Factor loading refers to the correlation between a variable and a factor. A high loading suggests that the variance explained by a variable is sufficient for it to have a considerable relationship with the factor. Small loadings (loadings $<0.3$) are considered insignificant~\cite{Costello2005Best,Hair2013Multivariate,Howard2015Review,Samuels2017Advice} and are thus suppressed in our model. For example, \textit{Cohesion.LCOM, Cohesion.LCOMModified, Cohesion.YALCOM,} and \textit{Size.CountInstanceVariable} load on together on Factor 1 which means that these variables seem to be measuring the same factor.

These factors explain 84\% of the variance in our dataset, which is good. If the variance explained was less than 60\%, we might opt to include additional factors~\cite{Hair2013Multivariate}. However, Factor 5 only has two metrics loading on it. The minimum is three, so either we need more metrics or fewer factors. 

In this case, we can reduce the number of factors to six, as theorized. Table \ref{tab:efa6f} shows the six-factor solution. This solution explains 78\% of the variance and each factor has at least six metrics, so we can move on to iteratively refining and interpreting the factors.

(We included this step to illustrate the realistic complexity of choosing an appropriate number of factors. Sometimes you're well into the analysis before you figure out how many factors you should have.)

\subsection{Interpret and Refine Factors}

Rotating the factors helps us interpret them. In a rotated factor solution, the axes is rotated so that variables that load together are plotted closer together on the axis causing them to load highly on a single factor. This provides a simpler, less complex structure to interpret.

Factors can be rotated using orthogonal or oblique rotation. An oblique rotation is preferable when factors are assumed to be correlated to each and an orthogonal rotation is used otherwise~\cite{Field2017Discovering, Costello2005Best, Hair2013Multivariate}. Despite the popularity of orthogonal rotation (specifically varimax), you should prefer to use oblique rotation since factors are usually correlated to each other. Use orthogonal rotation only if you have a very good reason to believe that the factors are uncorrelated. We use \textit{oblimin rotation} (a type of oblique rotation) to rotate the axis in our model.

Now we inspect the solution for problems, and remove problems one at a time, beginning with the worst. There is no algorithm for this. ``Worst'' is subjective. We can only give examples of problems and describe their severity. We are looking for three basic kinds of problems:

\begin{enumerate}
    \item Low communality: communality (h2) is the amount of variance in a variable that can be explained by the factor solution. Low communality (h2 $<$ 0.5) indicates that less than half of the variance of the variable is taken into account implying that the variable is not closely related to any of the factors and causes unwanted complexity with insufficient explanation~\cite{Hair2013Multivariate}. 
    \item Cross-loadings: variables with high loadings on multiple factors.
    \item Loading on the wrong factor: Variables loading highly (loadings $>$ 0.5) on a factor it shouldn't be loading highly on
\end{enumerate}

Looking at Table \ref{tab:efa6f}, \textit{Cohesion.LCOM5} has the lowest communality ($h2=0.16$) \textit{and} loads on the wrong factor (In-Coupling), so we remove that one first and re-run the EFA (Table~\ref{tab:efa6-1}). Now \textit{Size.CountDeclMethodDefault} has the lowest communality (h2=0.25) and loads on the wrong factor (In-Coupling again). So we remove it and run the EFA again (Table \ref{tab:efa6-2}). The next variable with lowest communality is \textit{Cohesion.YALCOM} (h2=0.45), however, it loads well on the correct factor so we'll retain it for now. 

Now that we've inspected all the obvious low communalities, we'll move on to crossloadings. \textit{Size.CountInstanceVariable} loads higher on the wrong factor (cohesion) than on the correct factor (size). Thus, we remove it from our analysis. Rerunning the EFA (Table \ref{tab:efa6-3}) we find that \textit{In-Coupling.CBOin} also has a cross-loading. However, it loads much higher on the correct factor (in-coupling) than the incorrect factor (out-coupling). Furthermore, the incorrect loading is smallest than the smallest correct loading in the EFA, so we retain \textit{Size.CountInstanceVariable} for now.  

Since there are no more low-communalities, cross-loadings or incorrect loadings, and the solution explains 87\% of the total variance in the dataset using six factors, each of which has at least three metrics, our EFA is now complete.

\renewcommand{\tabcolsep}{4pt}
\begin{table}[!htp]\centering
\sisetup{round-mode=places, round-precision=2}
\caption{EFA with 7 factors}\label{tab:efa7f}
\begin{tabular}{lccccccccc}\toprule
&\textbf{F1} &\textbf{F2} &\textbf{F3} &\textbf{F4} &\textbf{F5} &\textbf{F6} &\textbf{F7} &\textbf{h2} \\\midrule
Cohesion.LCOM &0.95 & & & & & & &0.92 \\
Cohesion.LCOM5 & & & & &0.64 & & &0.4 \\
Cohesion.LCOMModified &0.98 & & & & & & &0.93 \\
Cohesion.YALCOM &0.63 & & & & & & &0.45 \\
In-Coupling.CBOin & &0.87 &0.36 & & & & &1 \\
In-Coupling.FANINa & &0.93 & & & & & &0.94 \\
In-Coupling.FANINb & &0.99 & & & & & &1 \\
Out-Coupling.CBOout & & &0.8 & & & & &0.81 \\
Out-Coupling.FANOUTa & & &0.92 & & & & &0.9 \\
Out-Coupling.FANOUTb & & &0.98 & & & & &0.95 \\
Size.CountDeclMethodDefault & & & & &0.92 & & &0.9 \\
Size.CountInstanceVariable &0.44 & & &0.33 & & & &0.59 \\
Size.LOC & & &0.31 &0.71 & & & &0.69 \\
Size.NOM.Designite & & & &0.97 & & & &1 \\
Size.NOPM.Understand & & & &0.71 & & & &0.73 \\
Sub-Inheritance.CountSub & & & & & &1 & &1 \\
Sub-Inheritance.NC & & & & & &0.73 & &0.62 \\
Sub-Inheritance.SpecializationRatio & & & & & &0.94 & &0.9 \\
Sup-Inheritance.CountSup & & & & & & &0.98 &0.96 \\
Sup-Inheritance.DIT & & & & & & &0.9 &0.86 \\
Sup-Inheritance.ReuseRatio & & & & & & &0.96 &0.92 \\
\bottomrule
\end{tabular}
\end{table}
\begin{landscape}
\begin{table*}[!htp]\centering
\sisetup{round-mode=places, round-precision=2}
\caption{EFA with 6 Factors (Step 1)}\label{tab:efa6f}
\sisetup{detect-all=true}
\begin{tabular}{lcccccccc}\toprule
&\textbf{Cohesion} &\textbf{InCoupling} &\textbf{OutCoupling} &\textbf{Size} &\textbf{SubInheritance} &\textbf{SupInheritance} &\textbf{h2} \\\midrule
Cohesion.LCOM &0.93 & & & & & &0.92 \\
\textbf{Cohesion.LCOM5} & &\bfseries 0.42 & & & & &\bfseries 0.16 \\
Cohesion.LCOMModified &0.98 & & & & & &0.94 \\
Cohesion.YALCOM &0.62 & & & & & &0.45 \\
In-Coupling.CBOin & &0.84 &0.35 & & & &0.95 \\
In-Coupling.FANINa & &0.94 & & & & &0.92 \\
In-Coupling.FANINb & &0.96 & & & & &0.93 \\
Out-Coupling.CBOout & & &0.8 & & & &0.81 \\
Out-Coupling.FANOUTa & & &0.93 & & & &0.9 \\
Out-Coupling.FANOUTb & & &0.98 & & & &0.95 \\
Size.CountDeclMethodDefault & &0.53 & & & & &0.32 \\
Size.CountInstanceVariable &0.42 & & &0.34 & & &0.59 \\
Size.LOC & & &0.3 &0.74 & & &0.71 \\
Size.NOM.Designite & & & &0.9 & & &0.89 \\
Size.NOPM.Understand & & & &0.76 & & &0.71 \\
Sub-Inheritance.CountSub & & & & &1 & &1 \\
Sub-Inheritance.NC & & & & &0.73 & &0.63 \\
Sub-Inheritance.SpecializationRatio & & & & &0.94 & &0.91 \\
Sup-Inheritance.CountSup & & & & & &0.98 &0.96 \\
Sup-Inheritance.DIT & & & & & &0.91 &0.87 \\
Sup-Inheritance.ReuseRatio & & & & & &0.95 &0.91 \\
\bottomrule
\end{tabular}
\end{table*}
\end{landscape}
\begin{landscape}
\begin{table*}[!htp]\centering
\sisetup{round-mode=places, round-precision=2}
\sisetup{detect-all=true}
\caption{EFA with 6 factors (Step 2)}\label{tab:efa6-1}
\begin{tabular}{lcccccccc}\toprule
&\textbf{Cohesion} &\textbf{InCoupling} &\textbf{OutCoupling} &\textbf{Size} &\textbf{SubInheritance} &\textbf{SupInheritance} &\textbf{h2}  \\\midrule
Cohesion.LCOM &0.93 & & & & & &0.92  \\
Cohesion.LCOMModified &0.97 & & & & & &0.93 \\
Cohesion.YALCOM &0.62 & & & & & &0.45  \\
In-Coupling.CBOin & &0.87 &0.36 & & & &0.99  \\
In-Coupling.FANINa & &0.96 & & & & &0.95  \\
In-Coupling.FANINb & &0.99 & & & & &0.98  \\
Out-Coupling.CBOout & & &0.8 & & & &0.81  \\
Out-Coupling.FANOUTa & & &0.92 & & & &0.9  \\
Out-Coupling.FANOUTb & & &0.98 & & & &0.95 \\
\textbf{Size.CountDeclMethodDefault} & &\bfseries 0.43 & & & & &\bfseries 0.25  \\
Size.CountInstanceVariable &0.43 & & &0.35 & & &0.59  \\
Size.LOC & & & &0.73 & & &0.7 \\
Size.NOM & & & &0.96 & & &0.96  \\
Size.NOPM & & & &0.71 & & &0.67  \\
Sub-Inheritance.CountSub & & & & &1 & &1 \\
Sub-Inheritance.NC & & & & &0.73 & &0.62  \\
Sub-Inheritance.SpecializationRatio & & & & &0.94 & &0.9  \\
Sup-Inheritance.CountSup & & & & & &0.98 &0.96  \\
Sup-Inheritance.DIT & & & & & &0.91 &0.87  \\
Sup-Inheritance.ReuseRatio & & & & & &0.95 &0.91 \\
\bottomrule
\end{tabular}
\end{table*}
\end{landscape}

\begin{landscape}
\begin{table*}[!htp]\centering
\sisetup{round-mode=places, round-precision=2}
\sisetup{detect-all=true}
\caption{EFA with 6 factors (Step 3)}\label{tab:efa6-2}
\begin{tabular}{lcccccccc}\toprule
&\textbf{Cohesion} &\textbf{InCoupling} &\textbf{OutCoupling} &\textbf{Size} &\textbf{SubInheritance} &\textbf{SupInheritance} &\textbf{h2} \\\midrule
Cohesion.LCOM &0.94 & & & & & &0.92 \\
Cohesion.LCOMModified &0.98 & & & & & &0.95 \\
Cohesion.YALCOM &0.62 & & & & & &0.45 \\
In-Coupling.CBOin & &0.87 &0.36 & & & &1 \\
In-Coupling.FANINa & &0.95 & & & & &0.93 \\
In-Coupling.FANINb & &1 & & & & &1 \\
Out-Coupling.CBOout & & &0.8 & & & &0.81 \\
Out-Coupling.FANOUTa & & &0.92 & & & &0.9 \\
Out-Coupling.FANOUTb & & &0.98 & & & &0.95 \\
\textbf{Size.CountInstanceVariable} &\bfseries 0.42 & & &\bfseries 0.35 & & &\bfseries 0.59 \\
Size.LOC & & & &0.74 & & &0.7 \\
Size.NOM.Designite & & & &0.95 & & &0.93 \\
Size.NOPM.Understand & & & &0.71 & & &0.67 \\
Sub-Inheritance.CountSub & & & & &1 & &1 \\
Sub-Inheritance.NC & & & & &0.73 & &0.62 \\
Sub-Inheritance.SpecializationRatio & & & & &0.94 & &0.9 \\
Sup-Inheritance.CountSup & & & & & &0.98 &0.96 \\
Sup-Inheritance.DIT & & & & & &0.91 &0.86 \\
Sup-Inheritance.ReuseRatio & & & & & &0.96 &0.92 \\
\bottomrule
\end{tabular}
\end{table*}
\end{landscape}
\begin{landscape}
\begin{table*}[!htp]\centering
\sisetup{round-mode=places, round-precision=2}
\sisetup{detect-all=true}
\caption{EFA with 6 Factors (Step 4)}\label{tab:efa6-3}
\begin{tabular}{lcccccccc}\toprule
&\textbf{Cohesion} &\textbf{InCoupling} &\textbf{OutCoupling} &\textbf{Size} &\textbf{SubInheritance} &\textbf{SupInheritance} &\textbf{h2} \\\midrule
Cohesion.LCOM &0.95 & & & & & &0.95 \\
Cohesion.LCOMModified &0.96 & & & & & &0.91 \\
Cohesion.YALCOM &0.64 & & & & & &0.47 \\
\textbf{In-Coupling.CBOin} & &\bfseries 0.87 &\bfseries 0.36 & & & &\bfseries 1 \\
In-Coupling.FANINa & &0.95 & & & & &0.94 \\
In-Coupling.FANINb & &1 & & & & &1 \\
Out-Coupling.CBOout & & &0.8 & & & &0.81 \\
Out-Coupling.FANOUTa & & &0.92 & & & &0.9 \\
Out-Coupling.FANOUTb & & &0.98 & & & &0.95 \\
Size.LOC & & & &0.74 & & &0.72 \\
Size.NOM.Designite & & & &0.97 & & &0.98 \\
Size.NOPM.Understand & & & &0.67 & & &0.61 \\
Sub-Inheritance.CountSub & & & & &1 & &1 \\
Sub-Inheritance.NC & & & & &0.73 & &0.62 \\
Sub-Inheritance.SpecializationRatio & & & & &0.94 & &0.9 \\
Sup-Inheritance.CountSup & & & & & &0.98 &0.97 \\
Sup-Inheritance.DIT & & & & & &0.91 &0.86 \\
Sup-Inheritance.ReuseRatio & & & & & &0.96 &0.92 \\
\bottomrule
\end{tabular}
\end{table*}
\end{landscape}
\begin{landscape}
\begin{table*}[!htp]\centering
\sisetup{round-mode=places, round-precision=2}
\caption{Final EFA}\label{tab:efaFinal}
\begin{tabular}{lcccccccc}\toprule
&\textbf{Cohesion} &\textbf{InCoupling} &\textbf{OutCoupling} &\textbf{Size} &\textbf{SubInheritance} &\textbf{SupInheritance} &\textbf{h2} \\\midrule
Cohesion.LCOM &0.95 & & & & & &0.95 \\
Cohesion.LCOMModified &0.96 & & & & & &0.91 \\
Cohesion.YALCOM &0.64 & & & & & &0.47 \\
In-Coupling.CBOin & &0.87 & & & & &1 \\
In-Coupling.FANINa & &0.95 & & & & &0.94 \\
In-Coupling.FANINb & &1 & & & & &1 \\
Out-Coupling.CBOout & & &0.8 & & & &0.81 \\
Out-Coupling.FANOUTa & & &0.92 & & & &0.9 \\
Out-Coupling.FANOUTb & & &0.98 & & & &0.95 \\
Size.LOC & & & &0.74 & & &0.72 \\
Size.NOM.Designite & & & &0.97 & & &0.98 \\
Size.NOPM.Understand & & & &0.67 & & &0.61 \\
Sub-Inheritance.CountSub & & & & &1 & &1 \\
Sub-Inheritance.NC & & & & &0.73 & &0.62 \\
Sub-Inheritance.SpecializationRatio & & & & &0.94 & &0.9 \\
Sup-Inheritance.CountSup & & & & & &0.98 &0.97 \\
Sup-Inheritance.DIT & & & & & &0.91 &0.86 \\
Sup-Inheritance.ReuseRatio & & & & & &0.96 &0.92 \\
\bottomrule
\end{tabular}
\end{table*}
\end{landscape}

\end{document}